\documentclass[journal,comsoc]{IEEEtran}
\usepackage{cite}
\usepackage{xcolor}

\ifCLASSINFOpdf
  \usepackage[pdftex]{graphicx}
\else

\fi

\usepackage{amsmath,amssymb,amsfonts}
\usepackage{algorithm}
\usepackage{algorithmic}
\usepackage{csquotes}
\usepackage{soul}
\usepackage{color}
 \sethlcolor{yellow} 

\usepackage{tabularx,booktabs}
\usepackage{url}
\usepackage{caption}
\usepackage{subcaption}

\usepackage{amsmath,amsfonts, amssymb,enumitem,mathtools}
\usepackage{xcolor}
\usepackage{soul}
\usepackage{cite}
\sethlcolor{yellow}
\renewcommand\hl[1]{#1} 

\usepackage{amsthm}
\usepackage{soul}

\newcommand{\argmax}{\arg\!\max}

\usepackage{bbm}
\usepackage{dsfont}
\def\BibTeX{{\rm B\kern-.05em{\sc i\kern-.025em b}\kern-.08em
    T\kern-.1667em\lower.7ex\hbox{E}\kern-.125emX}}

\begin{document}


\title{UAV-Assisted Space-Air-Ground Integrated Networks: A Technical Review of Recent Learning Algorithms}

\author{Atefeh~Hajijamali~Arani,
        Peng~Hu,~\IEEEmembership{Senior Member,~IEEE,}
        and~Yeying~Zhu
\thanks{Atefeh Hajijamali Arani and Yeying Zhu are with the Dept. of Statistics and Actuarial Science, University of Waterloo, ON N2L 3G1, Canada (e-mail: ahajijam@uwaterloo.ca; yeying.zhu@uwaterloo.ca).}
\thanks{Peng Hu is with Department of Electrical and Computer Engineering, University of Manitoba, and the Faculty of Mathematics, University of Waterloo, ON N2L 3G1, Canada. He was with Digital Technologies Research Center, National Research Council Canada. (Corresponding author email: Peng.Hu@umanitoba.ca)}
}

\maketitle





\begin{abstract}
Recent technological advancements in space, air, and ground components have made possible a new network paradigm called \enquote{space-air-ground integrated network} (SAGIN). Unmanned aerial vehicles (UAVs) play a key role in SAGINs. However, due to UAVs' high dynamics and complexity, real-world deployment of a SAGIN becomes a significant barrier to realizing such SAGINs. UAVs are expected to meet key performance requirements with limited maneuverability and resources with space and terrestrial components. Therefore, employing UAVs in various usage scenarios requires well-designed planning in algorithmic approaches. This paper provides an essential review and analysis of recent learning algorithms in a UAV-assisted SAGIN. We consider possible reward functions and discuss the state-of-the-art algorithms for optimizing the reward functions, including Q-learning, deep Q-learning, multi-armed bandit, particle swarm optimization, and satisfaction-based learning algorithms. Unlike other survey papers, we focus on the methodological perspective of the optimization problem, applicable to various missions on a SAGIN. We consider real-world configurations and the $2$-dimensional ($2$D) and $3$-dimensional ($3$D) UAV trajectories to reflect deployment cases. Our simulations suggest the 3D satisfaction-based learning algorithm outperforms other approaches in most cases. With open challenges discussed at the end, we aim to provide design and deployment guidelines for UAV-assisted SAGINs.
\end{abstract}

\begin{IEEEkeywords}
Unmanned aerial vehicles, satellite networks, terrestrial networks, deployment, reinforcement learning, heuristic algorithms
\end{IEEEkeywords}


\maketitle

\section{Introduction}
\begin{figure*}[tb!]
\centering{\includegraphics[width=0.75\linewidth]{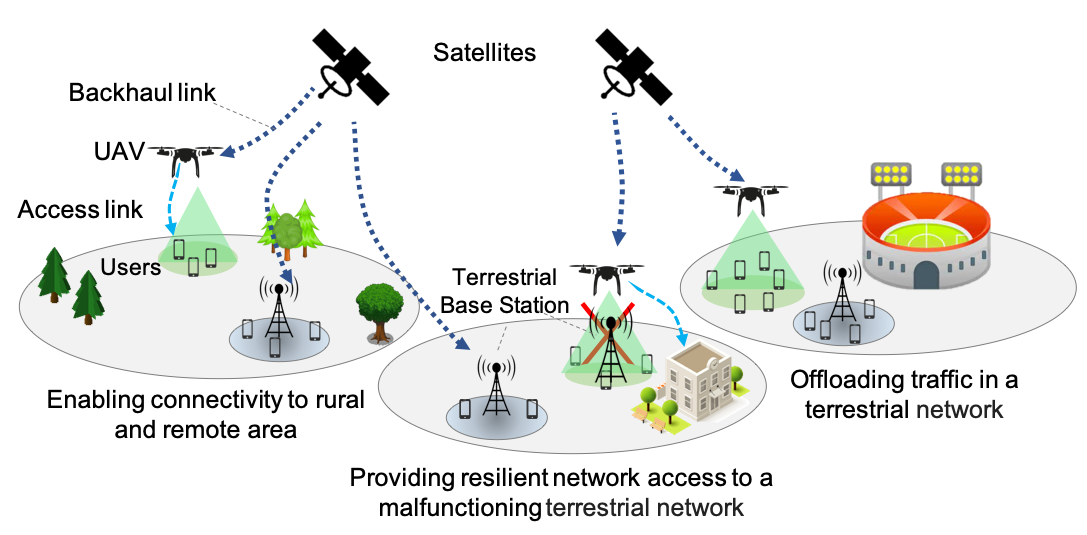}}
\caption{Example use cases of employing a UAV-assisted SAGIN for enabling network access to terrestrial network users, who are located in rural/remote areas and malfunctioning/overloaded terrestrial networks. UAVs in these cases are considered as aerial base stations. } \label{layout}
\end{figure*}

\IEEEPARstart{T}{he} recent advancements in the non-geostationary-orbit (NGSO) satellite networks, aerial and terrestrial networks have enabled the new paradigm called space-air-ground integrated networks (SAGINs). Unmanned aerial vehicles (UAVs) have great maneuverability and can significantly enhance the SAGIN's performance and resilience with well-designed planning. In a UAV-assisted SAGIN, UAVs play a critical role in the performance and resilience assurance by providing connectivity to users. In the representative scenarios depicted in Fig. \ref{layout}, UAVs can provide on-demand network access for ground users in unserved and underserved areas or adverse and overloaded conditions in a SAGIN-based network architecture. These scenarios can be extended to various setups and applications where UAVs play a role as an aerial base station (BS), where UAVs can also be viewed as a general high/low altitude platform stations (HAPSs/LAPSs) system \cite{Lin21} to enhance the coverage of satellite spot beams, which are subject to obstructions by rains, clouds, or other atmospheric conditions. UAVs can mitigate connection interruptions or outages caused by problematic terrestrial BSs \cite{Sun19, Zhong19, atefeh_access_2021}. In a generic scenario to offload high data traffic on a terrestrial network (TN) \cite{atefeh_PIMRC2020, atefeh_IoT2021, Fan21, Liao21}, UAVs can be dispatched to complete various tasks for providing ground users with consistent quality-of-experience (QoE).

However, the great promises of UAV-assisted SAGINs come with real-world challenges. First, for example, the modeling of the satellite networks, UAVs, and TNs needs to be made by key QoE requirements, such as throughput, network outage, and fairness.
Second, the use of network resources in all network segments needs to be jointly optimized. Third, the deployment of a UAV fleet needs to consider real-world factors, such as {energy consumption},
altitude keeping and trajectory planning, which can affect the problem modeling and performance \cite{shin_iot_2022, cai_MLBDBI22}. These challenges have not been systematically addressed in the literature. Recent survey papers as shown in Table \ref{sym_Tab} do not cover all topics for a SAGIN system model, such as UAVs, satellite components, ground components, and problem formulation and technical comparison. For example, authors in \cite{luong_applications_2019} discussed the overall use of reinforcement learning (RL) in communication networks, where the essential elements required in UAV-assisted SAGIN, such as satellite communication, ground components, problem formulation, and technical evaluation and comparison, are not addressed. In \cite{zhang_survey_2020}, the generic architectures using SAGINs in 5G networks are discussed without providing a consistent formulation and evaluation of problems with the use of UAVs and ground components. In \cite{Liu18}, the generic deployment overview of SAGINs is made, but no technical comparisons are made. Furthermore, only a portion of these papers discusses QoE metrics, although the metrics are application-specific. More importantly, these works have not systematically discussed the recent learning-based algorithmic approaches used in UAV-assisted SAGINs.

In recent years, both learning and non-learning-based methods have been proposed in the literature for optimally planning UAVs in various missions on a SAGIN. Most state-of-the-art works are focused on the use of RL approaches, while heuristic approaches are not included in the discussion. There are some research efforts addressing UAV network challenges from the non-learning perspective, such as in \cite{MAhdi2018comm,RZhang_2017_mount, RZhang_2018_TWC, Jiang_Xu_19, WangWei21,Cui19,JiangXu20}, which are 
developed based on successive convex approximation, penalty-based algorithms, and spatial average throughput for general and single UAV scenarios.
In the same context, the authors in \cite{Nguyen_gcom19} model the problem of two-dimensional ($2$D) placement of UAVs and channel allocation as a non-convex problem which is decoupled into two sub-problems.
To solve the problem, the difference between convex functions optimization and quadratic transformation techniques is adopted.
It is assumed that ground users served by UAVs are connected to the core network through a ground BS. 
These non-learning-based solutions may not be tractable for very complex and dynamic environments with high numbers of users and multiple UAVs. In addition, they require some prior knowledge about the system (e.g. the locations of users) which is impractical for real-time solutions. In this regard,  learning algorithms can assist in solving problems iteratively through learning from the system with low complexity and without the need for the full prior information of the system. Furthermore, the trajectories of UAVs are often modeled in a $2$D deployment scenario {which is addressed from a non-learning perspective such as Convex optimization \mbox{\cite{22}.} }
However, in a real-world setup, three-dimensional ($3$D) deployment is required to be considered. The $2$D and $3$D deployments change the trajectory planning of UAVs and have implications affecting various performance metrics. The evaluation of the applicable algorithmic approaches under $2$D and $3$D scenarios needs to be made. 
{Since traditional $2$D approaches are limited in capturing the complexity of $3$D spaces, some work based on deep RL (DRL) algorithms addresses this issue \mbox{\cite{Khoi2022, abohashish2023trajectory}}.} 
Some assumptions made at the users' level may be far from reality, such as statistic users and fixed user-BS associations. Thus, it is necessary to consider the user's mobility assumptions.

In order to examine learning-based methods for UAV-assisted SAGIN system designs, it is important to have an unbiased and up-to-date review of the major learning methods and to analyze these methods comparatively. The major contributions are summarized in the following:

\begin{itemize}
    \item We provide complete technical coverage of the UAV-assisted SAGIN as shown in Table \ref{Tbl:comp_surveys}. 
     \item We formulate the generic UAV-assisted SAGIN problem with implementations considering 2D and 3D UAV trajectory designs.
    \item We give an update-to-date discussion on applicable learning algorithms for UAV-assisted SAGIN, such as RL, DRL, satisfaction-based learning, and heuristic approaches. 
    \item We provide a consistent and systematic evaluation of the algorithmic approaches with real-world deployment considerations in essential QoE metrics.
   
\end{itemize}

{In our paper, we focus on optimizing UAV trajectories, placements, and resource allocations within SAGINs. These optimizations are pivotal for achieving desired performance metrics such as throughput, coverage, fairness, system load, and QoE. In this regard, we delve into various algorithms, including Q-learning, multi-armed bandit (MAB), deep Q-learning, heuristic methods (e.g., particle swarm optimization [PSO]), and satisfaction-based learning, and present their capabilities to address the challenges of integrating space, air, and ground components effectively. These algorithms offer innovative solutions, such as adaptive decision-making and practical, computationally efficient approaches. Furthermore, we conduct comprehensive simulations to compare these algorithms in realistic SAGIN deployment scenarios to highlight the relative strengths and limitations of each algorithm.
 Unlike some existing surveys, we focus on algorithmic approaches within UAV-assisted SAGINs. By examining these algorithms and their real-world deployment considerations, we provide invaluable insights and guidelines for researchers working on UAV-enabled integrated networks.
}
{The choice of RL and  PSO algorithms for SAGIN optimization is driven by the unique challenges of this integrated network paradigm. RL is selected for its adaptability to dynamic environments, complex decision-making capabilities, and ability to learn from experience, making it suitable for modeling real-world SAGIN scenarios. PSO is chosen for its population-based optimization, balancing global and local search, simplicity, and efficiency. These algorithms were specifically chosen for their applicability in addressing the complexities of integrating space, air, and ground components in SAGINs. The paper explores their applications in this context, highlighting their strengths and limitations in realistic SAGIN deployment scenarios.}

The remainder of the paper is structured as follows. Section \ref{sec_overview} reviews some recent developments of SAGINs. 
The formulation of a joint optimization problem is presented in Section \ref{sec_PF}.
Section \ref{sec_RL_SAGIN} overviews the RL approach and discusses the representative Q-learning and  MAB algorithms for SAGINs. Section \ref{sec_DRL} discusses the DRL approach for SAGINs. Section \ref{sec_satis} discusses the satisfaction-based learning approach for SAGINs. In Section \ref{sec_pso}, the PSO-based heuristic approach for SAGINs is discussed.  Evaluation of these algorithmic approaches and open challenges are discussed in Section \ref{sec_sim_result}. The conclusive remarks are made in Section \ref{sec_conclusion}.

{\textit{Notations:} The regular and boldface symbols refer to scalars and matrices, respectively. For any finite set ${\mathcal A}$, the cardinality of  set ${\mathcal A}$ 
is denoted by $\left|{\mathcal A}\right|$.
 The function $\mathbbm{1}_{\phi}$ denotes the indicator function which equals 1 if event $\phi$ is true and 0, otherwise.
$\ln(.)$ represents the natural logarithm function, which is the logarithm to the base $\mathrm{e}$.  The expression 
{$\mathrm{mod}(a, b)$} denotes the remainder of the division of 
$a$ by $b$.}

\begin{table*}[h]
\scriptsize
\centering{
\caption{{Recent works on algorithmic approaches for UAVs}} \label{sym_Tab}
\begin{tabular}{|p{0.35in}|p{0.3in}|p{0.6in}|p{0.6in}|p{0.6in}|p{0.6in}|p{0.5in}|p{0.3in}|p{0.3in}|p{0.6in}|p{0.3in}|p{0.3in}|} \hline 
{\textbf{Reference}}& \textbf{Year} &{\textbf{Addressed issues}}& \textbf{Access/backhaul links}& \textbf{Resource management} & \textbf{Trajectory design} & \textbf{Q-learning}& \textbf{MAB}& 
\textbf{Space}& \textbf{Aerial (single/multiple UAVs)} & \textbf{Ground}& \textbf{User mobility} \\ \hline

 \cite{atefeh_icc2021}& $2021$ & Throughput &Access, backhaul& Power, channel& $3$D  & {\textbf{\checkmark}}&{\textbf{\checkmark}} &
 \checkmark& Multiple &\checkmark &\checkmark \\ \hline
 
 {{\cite{atefeh_access_2021}}}& $2021$ & Load, throughput,  fairness & Access, backhaul&Channel& $3$D   &{\textbf{\checkmark}}&{\textbf{\checkmark}} &
 \checkmark& Multiple &\checkmark &\checkmark \\ \hline

 {{\cite{Nguyen_gcom19}}}& $2019$ & Throughput &  Access, backhaul&  Channel& $2$D   &{\textbf{-}}&{\textbf{-}} &
 -& Multiple &\checkmark& -\\ \hline
 
  {\cite{azade_GC19}}& $2019$ & Throughput & Access, backhaul & - & $2$D   & {{Non-learning}}&{Non-learning} &-
 & Multiple & \checkmark& \checkmark\\ \hline
 
  {{\cite{walid_joint2020}}}& $2020$ & Throughput & Access, backhaul &  Channel& -   &Non-learning&Non-learning & \checkmark & Multiple & \checkmark & -\\ \hline
 
   {{\cite{Cui_TWC20}}}& $2020$ & Energy efficiency & Access &  Power, channel& -  &\checkmark &{\textbf{-}} &-
 & Multiple & \checkmark& -\\ \hline

 {{\cite{Zdenek_vtc2020}}}& $2020$ & Throughput & Access, backhaul  &  Power& $2$D   &Non-learning&Non-learning & -
 & Multiple & \checkmark & - \\ \hline
 
  \cite{MAB_Emergency19}& $2019$ &  Battery consumption, throughput & Access &  -& $2$D   &-&\checkmark &
- & Single &- & -\\ \hline

  \cite{atefeh_PIMRC2020}& $2020$ & Throughput & Access &  -& $3$D   &\checkmark&\checkmark & - 
 & Multiple & \checkmark & -\\ \hline


\end{tabular}
}
\end{table*}

\begin{table*}
 \caption{Overview of Existing Survey Papers}
\label{Tbl:comp_surveys}
\renewcommand{\arraystretch}{1.05}
\begin{tabularx}{\linewidth}{
    >{\hsize=2\hsize}X
    >{\hsize=0.83\hsize}X
    >{\hsize=0.83\hsize}X
    >{\hsize=0.83\hsize}X
    >{\hsize=0.83\hsize}X
    >{\hsize=0.83\hsize}X
    >{\hsize=0.83\hsize}X
  }
\toprule
\textbf{Topic} & \textbf{References} & \textbf{Robots/UAVs } & \textbf{Satellite Components} & \textbf{Ground Components} & \textbf{Problem Formulation} & \textbf{Technical Comparisons} \\ 
\midrule
Survey on DRL in Communications Networks & \cite{luong_applications_2019} & Yes & No & No & No & No \\
\hline
Architectural overview on SAGIN in 5G networks & \cite{zhang_survey_2020} & No & Yes & No & No & No\\
\hline
Developments overview on SAGIN from network design, resource allocation, to
performance evaluation & \cite{Liu18} & Yes & Yes & No & No & No\\
\hline
Deep Reinforcement Learning for Internet of Drones Networks: Issues and Research Directions & \cite{noor_ojcs2023} & Yes & Yes & Yes & No& No\\
\hline
Our paper & {} & Yes & Yes & Yes & Yes & Yes \\
\bottomrule
\end{tabularx}
\end{table*}

\section{Overview of SAGINs} \label{sec_overview}
SAGIN is a recently proposed architecture \cite{Liu18} leveraging the space, aerial, and ground components. The space components may include geostationary (GEO), medium-Earth-orbit (MEO), and low-Earth-orbit (LEO) satellites \cite{kaddoum2023,park2023}. The aerial components may include HAPS/LAPS systems, such as stratospheric balloons, airships, and UAVs. The ground networks may include various telecommunications networks while a cellular network is the typical option used. Fig. \ref{layout} shows typical scenarios where satellites, UAVs, and ground networks are integrated into a SAGIN, where the essential links between satellites, UAVs, and cellular BSs are shown. Due to the breadth of the SAGIN topic, here we capture the key characteristics of a SAGIN and formulate a generic optimization problem considering UAV deployments and key QoE metrics, which can be extended to SAGIN variations.

SAGIN is a promising architecture that can address the recent developments in satellites and terrestrial networks and lead to 6G \cite{9520380}, but it also introduces many challenges from individual segments to an integrated system. From the aerial network perspective, a SAGIN can be assisted with UAVs being aerial BS nodes. Baltaci \textit{et al.} \cite{9508366} discussed the connectivity technologies and challenges based on satellite communication, HAPS networks, and air-to-air (A2A) links. 
Although UAVs, serving as aerial BSs, can provide better line-of-sight (LoS) coverage to ground users, their path planning needs to be optimized.
The placement of UAV-based BS has been explored in \cite{Lyu16}. A DRL is proposed in \cite{Wan19} to address the UAV path planning for mobile edge computing scenarios. The computation system poses another challenge where the computation resources from space, aerial, and ground components need to be made coherently. A scheduling scheme for computation offloading in SAGINs is proposed in \cite{9013393}. A cooperative scheme for utilizing the computation resources for different scenarios is proposed in \cite{9520380}. A deep Q-learning algorithm for traffic offloading is explored in \cite{Tang2022}. Heuristic methods represent another learning-based approach to solving UAV-related problems. For example, a PSO path planning scheme is explored in \cite{Shin20}. The adaptive PSO task scheduling scheme for a SAGIN is discussed in \cite{Dai19}. The UAV placement and coverage maximization problems using PSO are studied in \cite{Kalantari_2016, Yuheng_19}. We can see that both learning algorithms have recently been adopted in UAV and SAGIN settings. However, due to different setups and problem domains in the existing works, we can hardly see and compare actual performance in typical deployment scenarios shown in Fig. \ref{layout} between these algorithms. {Although} learning algorithms are promising to solve SAGIN-related problems, an overview from an algorithmic perspective is lacking in the current literature. When used as a toolbox for SAGIN research, it is essential to provide a systematic overview of these algorithms and discuss how they can be applied to the generic SAGIN system model.

\section{Problem Formulation} \label{sec_PF}
In this section, we discuss
the system model and problem formulation applicable to real-world scenarios.
The target is to maximize a predefined reward function adapted to the type of problem. 
In the context of providing connectivity to ground users, most existing literature focuses on maximizing throughput and coverage probability. For instance, the works in \cite{atefeh_icc2021, azade_GC19, walid_joint2020,Nguyen_gcom19,Zdenek_vtc2020} aim at maximizing the systems' data rates. However, these studies do not take into account the fairness among users, which is an important factor for evaluating the performance of a system. 

On the other hand, during high-traffic periods, ground BSs can become overloaded, resulting in degraded user throughput and increased inter-cell interference. An alternative and efficient complement to traditional TNs is UAV deployment, where the system configuration can be adapted to traffic demand and system load due to the flexibility of UAV placement.
Moreover, load balancing among UAV BSs needs to be optimized to achieve further improvement in spectral efficiency.   

{Here, we focus on backhaul communication in the system, presenting the architecture and parameters associated with backhaul links while considering millimeter-wave (mm-wave) communication.} The backhaul links comprise LEO satellites denoted as  $\mathcal L = \{1, \dots, |\mathcal L|\}$.  These satellites move in a circular orbit aligned with the y-axis at a fixed altitude $H_{\mathrm L}$ above Earth's surface. The orbital speed $V_{\mathrm{L}}$ is calculated using gravitational constants and Earth's mass parameters \cite{atefeh_access_2021}.
The total bandwidth $\omega_{\mathrm{BCK}}$ is allocated for backhaul communications and divided into $|\mathcal L|$ orthogonal channels, each with bandwidth $\omega_{\mathrm{BCK}}/|\mathcal L|$. The free space path loss  $L_{l,b}(t)$ between each satellite $l \in \mathcal L$ and each BS $b \in \mathcal B$ is calculated using the distance $d_{l,b}(t)$ at time $t$. Furthermore, the path loss $L_{l,b}(t)$ is determined based on the carrier frequency $f_c$, 
and is defined as follows \cite{Wang2024}:

\begin{equation}\label{path_loss}
    \begin{aligned}
L_{l,b}(t) = 20 \log_{10} (\frac{4\pi f_c d_{l,b}(t)}{c})  \  \ [\mathrm{dB}],
    \end{aligned}
\end{equation}
{where $c$ is the speed of light in a vacuum (approximately 299,792,458 m/s).}
In terms of distance in kilometers and frequency in MHz, \eqref{path_loss} can be {re-expressed} as \cite{Yan_survey_19, Salehi24, Xuchao24}
\begin{equation}\label{path_loss_v2}
    \begin{aligned}
L_{l,b}(t) = 32.45 + 20 \log_{10} (f_c) + 20 \log_{10} (d_{l,b}(t)) \  \ [\mathrm{dB}].
    \end{aligned}
\end{equation}

The channel gain $g_{l,b}(t)$ between satellite $l$ and BS $b$ at time $t$ is defined in \eqref{cha_gain_bck} which is determined by 
the path loss, $L_{l,b}(t)$, and antenna gains. Here, the transmit gain of satellite $l$'s antenna is denoted as $G_l^{\mathrm T}$, and the receive gain of BS $b$ is denoted as $G_b^{\mathrm R}$.
 Communication is established only if the distance $d_{l,b}(t)$ is within the maximum range $r_b^{\mathrm{max}}$.

\begin{equation}\label{cha_gain_bck}
  g_{l,b}(t) =  
  \begin{cases}
  10^{\frac{- L_{l,b}(t)}{10}} G_l^{\mathrm T} G_b^{\mathrm R}, & \mbox{if } d_{l,b}(t)\leq r_b^{\mathrm{max}}  \\
   0 , & \mbox{otherwise},
  \end{cases}
\end{equation}
where $r_b^{\mathrm{max}}$ represents the maximum height of a satellite above BS $b$'s horizon, enabling communication between the satellite and the BS.

Shannon's capacity formula calculates the achievable data rate $C_{l,b}(t)$ between BS $b$ and satellite $l$ at time $t$. This rate considers the association relation $a_{l,b}^{\mathrm{BCK}}(t)$, transmit power $p_l$, channel gain $g_{l,b}(t)$, and noise power $\sigma_0^2$. Hence, the achievable
data rate at BS $b$ associated with satellite $l$ at time $t$ is given
by

\begin{equation}\label{rate_bck}
\begin{aligned}
C_{l,b}(t)= \frac{\omega_{\mathrm{BCK}}}{|\mathcal L|} \log_2(1 + \frac{a_{l,b}^{\mathrm{BCK}}(t) \  p_l \  g_{l,b}(t)}{\sigma_0^2})    \  \ [\mathrm{bps}],
\end{aligned}
\end{equation} 
where the binary element $a_{l,b}^{\mathrm {BCK}}(t) \in \{0, 1\}$  denotes  the association relation between satellite $l$ and BS $b$. Specifically, $a_{l,b}^{\mathrm {BCK}}(t) = 1$ indicates that BS $b$ is associated to satellite $l$ at time $t$, otherwise $a_{l,b}^{\mathrm {BCK}}(t) = 0$. 

Now, we present the access link model. Let $\Psi_k$ and $\mathcal K_b$ denote the requested rate of user $k$ and the set of users associated with BS $b\in\mathcal B$, respectively, where $\mathcal B = \mathcal U  \cup \mathcal S$ is the set of all the BSs. Here, $\mathcal U$ and $\mathcal S$ are the set of UAVs and small cell BSs (SBSs), respectively. 
The load of BS $b\in\mathcal B$ is defined as \cite{atefeh_tvt}

\begin{equation}\label{load_coupled}
\begin{aligned}
&\rho_b  = \sum_{k\in\mathcal K_b} \frac{\Psi_k}{  ~ C_{b,k}}\triangleq f_b(\boldsymbol\rho),
\end{aligned}
\end{equation}
where $ C_{b,k}$ is the achievable data rate to user $k$ provided by BS $b$.
Here, function $ f_b(.)$  represents the load of BS $b$ as a function of the loads of all the BSs, where  $\boldsymbol \rho=\big(\rho_1, \dots, \rho_{|\mathcal B|}\big)$. To find the loads of the BSs, we use the fixed point iteration algorithm due to the fact that the function $f_b(\boldsymbol\rho)$ is a standard interference function. 

Now, we describe Jain's fairness index, one of the most widely-used fairness metrics,   which provides a quantitative assessment of fairness among users.   Jain's fairness index can be defined as follows \cite{jain1984quantitative}:

\begin{equation}\label{fairness_metric}
\mathcal F = \frac{\big(\sum_{k\in\mathcal K}\bar{C}_k \big)^2}{|\mathcal K| \big(\sum_{k\in\mathcal K}\bar{C}_k ^2\big)},
\end{equation}
where $\bar{C}_k$ and $\mathcal K$ are the total data rate for user $k$ and the set of users, respectively.

Here, we aim at maximizing the fairness while balancing load among UAVs flying over the particular area  $\mathcal R$.
Therefore, we define a reward function that captures the fairness among users  and the load of BSs  as follows:

\begin{equation}\label{reward_func}
\Gamma_b(t) = \Phi_{b} \mathcal F(t) + \psi_{b} (1-\rho_b(t)),
\end{equation}
%
where the coefficients $\Phi_{b} $ and $\psi_{b}$ are the weight parameters that indicate the impact of the fairness and load on the reward function, respectively. 
Our overall objective is to maximize the total system reward function by optimizing the trajectories of the UAVs and channel allocation at the BSs as given by the following optimization problem: 

\begin{subequations} \label{max_prob}
\begin{align}
\max _{\substack{\boldsymbol q(t), \boldsymbol A(t)}}
~~ & \sum_{t\in N}\sum_{b\in \mathcal B}  \Gamma_b(t) \\
\text { s.t. } &  (x_u(t),y_u(t))\in \mathcal R, \quad \forall u \in \mathcal U, \label{eqe1}\\
& h_u(t)\in[h_{\mathrm{min}},h_{\mathrm{max}}], \quad \forall u \in \mathcal U, \label{eqe2}\\
& \rho_b(t) = f_b(\boldsymbol\rho),~ ~ \forall b \in \mathcal B, \label{load_vct}\\
& 0\leq \rho_b(t)\leq 1, \quad \forall b \in \mathcal B \label{load_indivi},
\end{align}
\end{subequations}
where $\boldsymbol q(t)$ and $\boldsymbol A(t)$ are the vector of all the BSs' transmit channels and the locations of all the UAVs, respectively. $N$ is the total time instants. 
For the optimization problem \eqref{max_prob}, we consider several constraints.
The constraints  in \eqref{eqe1} and \eqref{eqe2}  define the feasible area for the locations of the UAVs in the $3$D space. The constraints in \eqref{load_vct} and \eqref{load_indivi} are related to the definition of load. Here, $\boldsymbol a_u(t) = (x_u(t), y_u(t), h_u(t))$  denotes the $3$D coordinate of UAV $u$ at time $t$. $h_{\mathrm{min}}$ and $h_{\mathrm{max}}$ are the minimum and maximum altitude of the UAVs.

\section{Reinforcement Learning for SAGINs} \label{sec_RL_SAGIN}

RL is a feedback-based machine learning (ML) technique in which agents learn to interact with the
environment by selecting actions and observing their outcomes \cite{kaelbling1996,sutton2018reinforcement}. Theoretically, RL algorithms use the Markov decision process (MDP) framework composed of an environment and a set of agents \cite{puterman2014markov}. 
Agents face a trade-off between \textit{exploration} and \textit{exploitation}, in which each agent exploits the action with the highest reward and explores its other actions to enhance the estimations of actions' rewards.
The main elements of RL algorithms can be defined as follows:

\begin{itemize}
    \item \textit{Agent}: A decision maker that can explore the environment to take an action, and receives a reward associated with its action. 
    \item \textit{Environment}:  A situation that an agent  operates and {is} surrounded by.

    \item \textit{Action}: 
    An action is a decision taken by an agent. 
    
    \item \textit{Reward}: 
    Feedback is feedback that an agent receives from the environment after taking action to evaluate its performance.
    
    \item\textit{State}: It is the current situation of an agent returned by the environment in effect of its selected action.
    
\item \textit{Value Function}: It indicates the expected return with a discount factor  for each state, given a certain policy.
  
\end{itemize}

{Now, we elaborate on the classification of learning algorithms which plays a crucial role in understanding their applicability and effectiveness, particularly within SAGINs. Learning algorithms in our work are categorized based on key characteristics and methodologies, including model-based vs. model-free, value-based vs. policy-based, and on-policy vs. off-policy algorithms.}

\begin{itemize}
   \item { \textbf{Model-based vs. model-free algorithms}}
    
\begin{itemize}
    \item 
{Model-based algorithms:} These algorithms utilize an explicit environment model to make decisions and learn optimal policies. They often involve building a predictive model of the system dynamics to estimate future states and outcomes.
    \item 
{Model-free algorithms:} In contrast, model-free algorithms directly learn optimal policies or value functions from interaction with the environment without explicitly modeling its dynamics. They rely on trial-and-error learning and do not require prior knowledge of the system dynamics.
\end{itemize}

 \item {\textbf{Value-based vs. policy-based algorithms}}
 \begin{itemize}
     \item 
{Value-based algorithms:} These algorithms aim to learn the value function associated with different actions or states in the environment. They focus on estimating the expected return or utility of taking certain actions and select actions based on maximizing this value function.
 \item 
{{
Policy-based algorithms:} } Policy-based algorithms directly learn the policy or behavior function that maps states to actions. Instead of estimating the value of each action, they aim to find the optimal policy directly by optimizing the policy parameters.
 \end{itemize}

 \item 
{\textbf{On-policy vs. off-policy algorithms:}}
\begin{itemize}
    \item {On-policy algorithms:} On-policy algorithms update their policies based on the data collected from the current policy. They directly optimize the policy that is currently being followed, which can lead to more stable learning but may suffer from slow convergence.
 \item {Off-policy algorithms:} Off-policy algorithms learn from data generated by a different (often exploratory) policy than the one being optimized. They can potentially learn more efficiently from diverse data sources but may require techniques such as importance sampling to account for the mismatch between the data distribution and the target policy.
\end{itemize}
\end{itemize}

Among different RL algorithms, we focus on MAB and Q-learning algorithms which are mostly used in literature to solve problems in various {applications}. {To be noticed, recent literature has demonstrated the effectiveness of deep deterministic policy gradient (DDPG)  and proximal policy optimization (PPO) in similar networking scenarios, highlighting their capability to handle complex decision-making tasks and optimize system performance\mbox{\cite{DDPG1, DDPG2}}. However, these two RL algorithms are designed to handle continuous action spaces. In DDPG, the actor-network outputs continuous action values directly, allowing for fine-grained control in environments with continuous action spaces. Similarly, PPO is capable of handling continuous action spaces through techniques such as parameterizing the policy with a neural network and optimizing it using stochastic gradient ascent. Therefore, both DDPG and PPO are tailored for continuous action spaces, making them valuable tools for tasks where actions are not restricted to discrete choices. As in our work, we mainly focus on discrete action spaces instead of continuous ones, we do not incorporate the particular comparison for these two RL approaches, but it certainly leaves room for future research. }

\subsection{Q-learning Algorithm}

\begin{figure}[b!]
\centering
{\includegraphics[width=0.7\linewidth]
{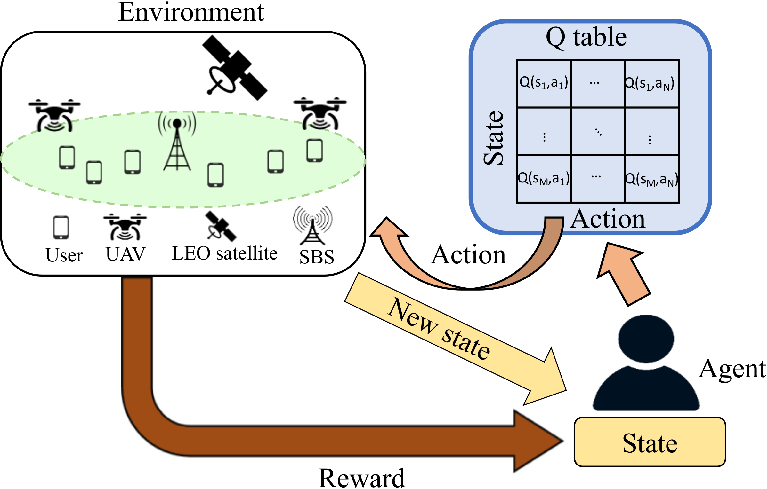}}
\caption{Learning structure based on the Q-learning algorithm.} \label{Qlearn_str}
\end{figure}

Q-learning algorithm is known as one of the most popular algorithms among RL algorithms\cite{kaelbling1996,watkins1992q}. In this regard, we intend to provide a review on the applications of Q-learning algorithms in UAV-enabled systems. 
{For a better understanding of its application, we first present the fundamentals of the Q-learning algorithm.}
{Q-learning is a fundamental RL algorithm that learns to maximize cumulative rewards by iteratively updating the Q-values (action values) of state-action pairs based on observed rewards and transitions.
Q-learning is categorized as a model-free, value-based, and off-policy RL algorithm. It is model-free because it does not require a model of the environment's dynamics. It is value-based because it estimates the value of a particular action in a given state. It is off-policy because it learns from a behavior policy different from the target policy.}

Let $\pi$ denote a \textit{policy} for an agent that maps a state to an action. The goal is to find an optimal policy which maximizes the expected sum of discounted reward instead of the immediate reward. Here, we define the value function $\mathcal V_{\pi}(s,a)$ for policy $\pi$ and taking action $a$ in state $s$ which can be given as follows \cite{luong_applications_2019}:

\begin{equation}\label{value_fun}
\begin{split}
\mathcal V_{\pi}(s,a) = & \mathbbm{E}_{\pi}\Big[\sum_{t=0}^{\infty} \gamma_{\mathrm{QL}}r(t)|s_0 = s \Big] = \\
& \mathbbm{E}_{\pi}\big[r(t)+\gamma_{\mathrm{QL}} \mathcal V_{\pi}\big(s({t+1}),a({t+1})\big)|s_0 = s\big],
\end{split}
\end{equation}
where $\gamma_{\mathrm{QL}}$ and $r(t)$ are the discount factor and the reward at time $t$, respectively.
We can observe that the value function can be decomposed into two parts including the immediate reward and the discounted value of the successor state. Accordingly, we aim to obtain the optimal policy that maximizes the value function as

\begin{equation}\label{value_fun2}
\begin{split}
\mathcal V_{\pi}^*(s,a) = \max_{a_t}\Big\{\mathbbm{E}_{\pi}\big[r(t)+\gamma_{\mathrm{QL}} \mathcal V_{\pi}\big(s({t+1}),a({t+1})\big)\big]\Big\}.
\end{split}
\end{equation}
Let $Q^*_{\pi}(s,a) \triangleq r(t)+\gamma_{\mathrm{QL}}\mathbbm{E}_{\pi}\big[\mathcal V_\pi\big(s({t+1}),a({t+1})\big)\big]$ be the optimal Q-function. Thus, the optimal value function can be represented by $\mathcal V^*(s,a)=\max_{\pi} Q^*_{\pi}(s,a)$. To find the optimal values of Q-function $Q^*_{\pi}(s, a)$, an iterative process can be used. Therefore, the
Q-function can be updated as follows:

\begin{equation}\label{Q-learning_eq}
\begin{aligned}
Q\big(s(t),a(t)\big)\leftarrow & Q\big(s(t),a(t)\big)+ \alpha_{\mathrm{QL}}(t) \big[r(t)+ \gamma_{\mathrm{QL}}\\
& \max_{a^\prime}Q\big(s({t+1}),a^\prime\big)
-Q\big(s(t),a(t)\big)\big].
\end{aligned}
\end{equation}
 The update rule in \eqref{Q-learning_eq}  finds the Temporal
Difference (TD) between the predicted Q-value $r(t)+ \gamma_{\mathrm{QL}} \max_{a^\prime}Q\big(s({t+1}),a^\prime\big)
-Q\big(s(t),a(t)\big)$ and the current value $Q\big(s(t),a(t)\big)$. Here, $Q\big(s(t), a(t)\big)$ is the Q-function (or the learned action-value function) for taking action $a(t)$ in state $s(t)$ at time $t$. Parameter $\alpha_{\mathrm{QL}}(t)$ denotes the learning rate which shows the impact of new
information to the existing value, and it is chosen according to the following conditions: $\alpha_{\mathrm{QL}}(t)\in[0,1]$, $\lim_{t\rightarrow\infty}\sum_{t=0}^\infty \alpha_{\mathrm{QL}}(t) = +\infty$, and $\lim_{t\rightarrow\infty}\sum_{t=0}^\infty \big(\alpha_{\mathrm{QL}}(t)\big)^2<\infty$.

As illustrated in Fig. \ref{Qlearn_str}, the Q-learning algorithm is based on a Q-table for each agent at each step. 
Let $\mathcal S$ and   $\mathcal A$ denote the set of all possible
states and actions, respectively.
The table consists of the combinations of states and actions, in which the dimension of the table is $|\mathcal S|\times|\mathcal A|$.
An example of the state and action of an agent in a path planning problem can be the location of the agent in a given area and its movement in different directions, respectively  \cite{sousa_19_pimrc,atefeh_access_2021}. 
Algorithm \ref{alg_qlearn} shows the pseudocode for the Q-learning algorithm in a multi-agent system with  $|\mathcal B|$ agents where $\mathcal B$ is the set of agents.  {In our context, each agent in the system is represented by a UAV.}
The subscript $b$ represents the elements of the RL  algorithm for agent $b$, e.g., $a_b(t)$ denotes the action of agent $b$ and $s_b(t)$ is the state of agent $b$ at time $t$.
Each agent $b$ interacts with the environment, and selects action $a_b(t)\in\mathcal A_b$ {at time } {{$t\in\{1, \dots, N\}$}  with duration \mbox{$T_s$}, where {$\mathcal A_b$} is the set of actions for agent $b$,  and {$N$} is the total number of  time instants. 
Then, it transits from state $s_b(t)$ to a new state $s_b({t+1})$, and receives the reward  $r_b({t})$. Furthermore, an iterative Q-function is updated under a stochastic state and the action taken by the agent using a learning rate $\alpha_{\mathrm{QL}}(t)$ and a discount factor $\gamma_{\mathrm{QL}}$ as described in \eqref{Q-learning_eq}.
%
}
The learning rate  $\alpha_{\mathrm{QL}}(t)\in[0,1]$ indicates the impact of the old value of the action-value function on the current update. Parameter $\gamma_{\mathrm{QL}}$ determines the impact of the future reward on the system and balances  the importance of short-term and long-term reward
%
which is in the range  $[0, 1]$. 
Note that, by allowing $\gamma_{\mathrm{QL}}\rightarrow 0$ it leads to considering the immediate reward of an action.  On the contrary, by allowing $\gamma_{\mathrm{QL}}\rightarrow 1$, the future reward has the same weight as the immediate reward. 
After updating the Q-function, the agent selects the action with the highest Q-value with probability $1-\epsilon$, and with probability $\epsilon$, it selects a random action to ensure exploration of the state-action space.
%
Let us discuss the time complexity of the Q-learning algorithm in  Algorithm  \ref{alg_qlearn}. If we denote 
$|\mathcal S|$ as the size of the state space, the worst-case complexity for action executions in steps 4-12 is bounded by  
$O(|\mathcal S|^3)$
\cite{Koenig-1993-15970}. Therefore, the time complexity of  Algorithm  \ref{alg_qlearn} is
$O( N \cdot |\mathcal S|^3)$.



\begin{algorithm}[t]
\caption{: Q-learning Algorithm}
\label{alg_qlearn}
\begin{algorithmic}[1]

\STATE \textbf{Input}: $\mathcal{B}$, $\mathcal S$, $\mathcal A_b$,  $\forall b \in \mathcal B$, $N$, \(\alpha_{\mathrm{QL}}(t)\) for \(t = 0\), $\gamma_{\mathrm{QL}}$, $\epsilon$

\STATE \textbf{Outputs}: $a_b(t), Q(s_b(t), a_b(t))$  $\forall b \in \mathcal B$, $ t \in \{1, \dots, N\} $
\STATE \textbf{Initialization}:  
$Q(s_b(t), a_b(t)) = 0$ for all $s_b(t) \in \mathcal S$ and $b\in\mathcal{B}$ for $t = 0$,

\WHILE{$t<N$}
\STATE $t \leftarrow t+1$
\FOR{$\forall b \in \mathcal B$}
\IF{$rand(.) < \epsilon$}
\STATE Select action $a_b(t)$ randomly
\ELSE
\STATE Select action $a_b(t) = \argmax_{a^\prime_b} Q\big(s_b(t),a^\prime_b\big)$
\ENDIF
\STATE Calculate reward $r_b({t})$, and observe the state $s_b({t+1})$
\STATE Update Q-function according to \eqref{Q-learning_eq}
\ENDFOR
\ENDWHILE
\end{algorithmic}
\end{algorithm}%


Learning algorithms can play an essential role in enhancing the performance of integrated networks. Several studies explore the use of Q-learning algorithms for designing UAV trajectories and path planning to meet the QoE requirements and achieve system performance objectives, such as maximizing coverage and throughput, minimizing interference, and optimizing QoE.
In \cite{sousa_19_pimrc}, a trajectory optimization problem is studied through a RL algorithm which consists of using a Q-learning-based approach for a scenario where multiple UAVs aim at maximizing the sum rate of ground users.{ The UAVs are trained to determine their trajectories according to the system topology and try to decrease their distances to the users. This can result in enhancing the communication link quality. The reward function of each UAV captures three terms, including the sum rate of users associated with the UAV, the distance between the UAV and its final location, and an activation
function to assure the safety of the UAVs. It is also assumed that the altitudes of UAVs are fixed to a certain value. }
Finding the optimal trajectory of a single UAV flying at a constant altitude is studied in \cite{Gesbert18}. Using a Q-learning-based algorithm, the UAV learns its $2$D location to maximize the sum rate of ground users { in the environment which contains a cuboid obstacle with a
height equal to the altitude of the UAV. The reward function includes the sum rate and a negative component which avoids the UAV stepping outside the area. Moreover, an additional term is added to the reward function for the UAV safety check, in which it can return to its initial location within the flying time limit.  }
To maximize the sum rate while satisfying the rate requirement of users, a three-phase mechanism is developed in \cite{Hanzo19}. In the first phase,  a Q-learning algorithm is applied to determine the locations of  UAVs according to the initial locations of users. Then, using a real dataset, the trajectories of users are determined, and then their future positions are predicted.  Finally, a Q-learning scheme is proposed to find the transmit power and predict the locations of UAVs based on the mobility of users in the network. {The reward function of the Q-learning corresponds to the instantaneous sum
rate of the users.}
The authors in \cite{LiuXiao_19} investigate the placement of multiple UAVs to maximize the sum mean opinion score which evaluates the satisfaction of users. To solve the problem, a three-step approach is provided.  First, a genetic algorithm based {on} K-means is applied to find the cell partition of users,{ in which the users are partitioned into different clusters, and a UAV is deployed for each cluster.} Next, by using a Q-learning algorithm, the locations of UAVs are initially determined when users are static. Then, for the case that users are roaming, a Q-learning-based movement algorithm is developed and a discrete reward function is used based on the instantaneous mean opinion score of the users.
Authors in \cite{Cui_TWC20} investigate the problem of resource management in a multi-UAV network to efficiently allocate power and channels. The objective is to maximize
a reward function, which captures power and throughput, to improve the energy efficiency of the system through allocating power and channels. 
Furthermore, they consider predefined flight plans for UAVs.
In the cellular network context, the authors in \cite{Liu_JCIN19} address the trajectory design problem for a single UAV to maximize the satisfied users. The satisfactions of users are determined based on the completion of user's request within an endurance time.  Compared to other works, they consider two types of users including ground and aerial users. To solve the problem, a double Q-learning algorithm is used which yields an improvement over a Q-learning-based algorithm.

\subsection{MAB Algorithm}
\begin{figure}[b!]
\centering\resizebox{3.3in}{1.9in}{\includegraphics
{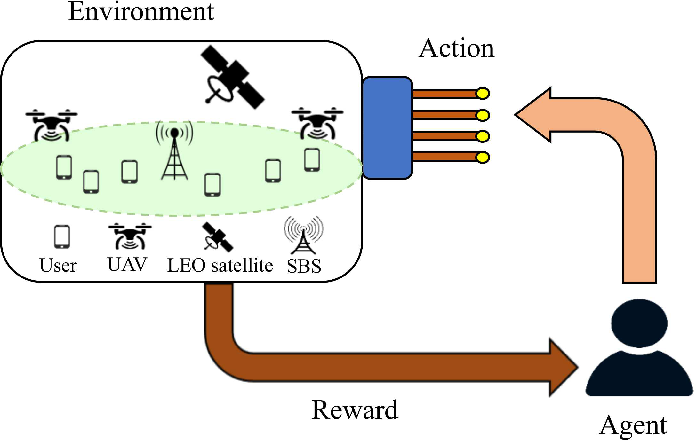}}
\caption{Learning structure based on MAB algorithm.} \label{MAB_str}
\end{figure}

In the MAB approach, each agent (or bandit) has multiple arms (or actions). After choosing an action from its set of actions, the agent observes the reward associated with the selected action and updates the average reward of that action accordingly \cite{katehakis1987multi}. However, this approach does not require any prior knowledge of the actions' rewards. Therefore, agents try to explore different actions. Fig. \ref{MAB_str} illustrates the structure of the MAB algorithm.
 {MAB algorithms can be either model-free or model-based, depending on the specific approach used. They are typically value-based as they involve estimating action values or policy-based if they maintain a probability distribution over actions. Since MAB algorithms update their policies based on current experiences, they are generally considered on-policy.}
One approach to solving MAB-based problems is the upper confidence bound (UCB) policy, where the action is chosen as \cite{samad_18_glob}:

\begin{equation}\label{MAB_reward}
\begin{split}
a_b^{\mathrm{MAB}}(t)  = \mathop{\argmax }_{a_{b,i} \in \mathcal A_b} \Big\{\Bar{ R}_{b,i}(t)+  \sqrt{\frac{2 \ln t}{n_{b,i}(t)}}\Big\},
\end{split}
\end{equation}
where $\mathcal A_b$ and  $\Bar{R}_{b,i}(t)$ represent the action set of agent $b$ and the average reward from action $a_{b,i} \in \mathcal A_b$ for player $b$ at time  $t$, respectively. Parameter $n_{b, i}(t)$ is the number of times that action $a_{b, i}$ has been selected by agent $b$ until time  $t$. {The pseudocode for the MAB algorithm is presented in Algorithm \mbox{\ref{alg_ucb}.}}
%
For the time complexity of the MAB algorithm described in Algorithm  \ref{alg_ucb}, if we consider the initialization step 1 takes $t_0$ and steps 5-13 take $t_1$, the total time taken is $t_0 + N\cdot|\mathcal{B}|\cdot t_1$. As $t_0$ is a constant, the time complexity of  Algorithm  \ref{alg_ucb} can be derived as $O(N \cdot |\mathcal B|)$.

\begin{algorithm}[t]
\caption{: MAB Algorithm}
\label{alg_ucb}
\begin{algorithmic}[1]
\STATE \textbf{Inputs}: $\mathcal{B}$, $\mathcal A_b$,  $\forall b \in \mathcal B$
\STATE \textbf{Outputs}: $\Bar{R}_{b,i}(t)$,  $a_b^{\mathrm{MAB}}(t)$, $\forall b \in \mathcal B$, $\forall a_{b,i} \in \mathcal A_b$, $ t \in \{1, \dots, N\} $

\STATE \textbf{Initialization}:  $\Bar{R}_{b,i}(t)=0$, $n_{b,i}(t)=0$ for $t=0$,  $\forall b \in \mathcal B$, $\forall a_{b,i} \in \mathcal A_b$ and $i\in\{1, \dots, |\mathcal A_b|\}$
\WHILE{$t<N$}
\STATE $t \leftarrow t+1$
\FOR{$\forall b \in \mathcal B$}
\IF{$\exists a_{b,i} \in \mathcal A_b$ s.t. $n_{b,i}(t)=0$}
\STATE{Select arm $a_b^{\mathrm{MAB}}(t) = a_{b,i}$}
\ELSE
\STATE{Select arm $a_b^{\mathrm{MAB}}(t)$} according to \eqref{MAB_reward}
\ENDIF
\STATE Calculate $ R_b(t)$ 
\FOR{$\forall a_{b,i}\in \mathcal A_b$}
  \STATE Update $a_{b,i}(t)$ as:  $n_{b,i}(t) =\!n_{b,i}(t-1) + \mathds{1}_{\{a_{b,i} = a_b^{\mathrm{MAB}}(t)\}}$
 \STATE Update $\Bar{R}_{b,i}(t)$ as: \\ 
 $\Bar{R}_{b,i}(t) = \frac{n_{b,i}(t-1) {\Bar{R}}_{b,i}(t-1)+\mathds{1}_{\{a_{b,i} = a_b^{\mathrm{MAB}}(t)\}} R_b(t)}{n_{b,i}(t)}$
\ENDFOR
\ENDFOR
\ENDWHILE
\end{algorithmic}
\end{algorithm}

{The main difference between the MAB and Q-learning algorithms is that the MAB algorithm does not recognize the state of the system and is only based on actions. In contrast, the Q-learning algorithm depends on the state of the system to update the Q-value function,  making it a state-action-based algorithm.}
This distinction is crucial when considering their applications in real-world scenarios. For instance, authors in \cite{atefeh_icc2021 } address the joint backhaul and access link optimization for a  SAGIN. The problem of satellite-BS association in backhaul links is solved to maximize throughput. In access links, the problem of BS-user association,  $3$D trajectory of UAVs, and resource management for {SBSs)} and UAVs are investigated to improve 
 system throughput. 
To solve the access link problem, a UCB-based mechanism is utilized.
On the other hand, the load of SBSs and UAVs is considered as the function of the user’s required rate to capture user heterogeneity. 
In the same context, the authors in \cite{atefeh_access_2021} leveraged the MAB mechanism to take into account provisioning fairness among users and balancing load among SBSs and UAVs. The proposed approach is compared to a Q-learning-based mechanism and yields better performance in terms of fairness, load, and throughput. 
%
A UAV-assisted emergency communication solution is developed in  \cite{MAB_Emergency19}. In this solution, the UAV acts as an aerial BS to provide communication services for ground users in a  post-disaster
area.  The target is to find the optimal path to serve the maximum number of users.  The UAV task is formulated as an extended MAB, and two solutions based on the distance-aware UCB and $\epsilon$-exploration algorithms are proposed.

In the context of anti-jamming strategy, the authors in  \cite{jamming_2020} propose a MAB-based anti-jamming channel selection model for
software-defined UAV swarm systems. In \cite{Combinatorial_MAB20}, a set of UAVs serve ground users as edge computing servers with the objective to minimize the delay of the offloaded tasks over time.  The task offloading problem is modeled as a combinatorial MAB problem, and a combinatorial bandit UCB algorithm is developed. In \cite{task_iot19_vtc}, a UAV task offloading problem based on MAB is addressed, followed by the development of a variance-reduced learning-aided task offloading scheme. In the context of non-orthogonal
multiple access (NOMA)-UAV systems, a MAB-based approach is used in \cite{NOMA_UAV21}, where a single UAV collects data from the Internet of Things (IoT) sensors in the uplink direction. The objective of the problem is to maximize the sum rate of all sensors.

\section{Deep Reinforcement Learning for SAGINs}\label{sec_DRL}
Although RL algorithms can be used to solve different types of complex decision-making problems in various fields, they yield degraded
 performance when state spaces and action sets are large.  Hence,  problems with large and/or continuous states can be difficult to solve with traditional RL methods. 
DRL algorithms have been shown to be a promising solution for tackling the limitations of RL \cite{Gao_GCN21}. DRL can be treated as a combination of RL and function approximation, where function approximation is used to approximate the Q-value function. 

\begin{figure}[tb!]
\centering{\includegraphics[width=\linewidth]
{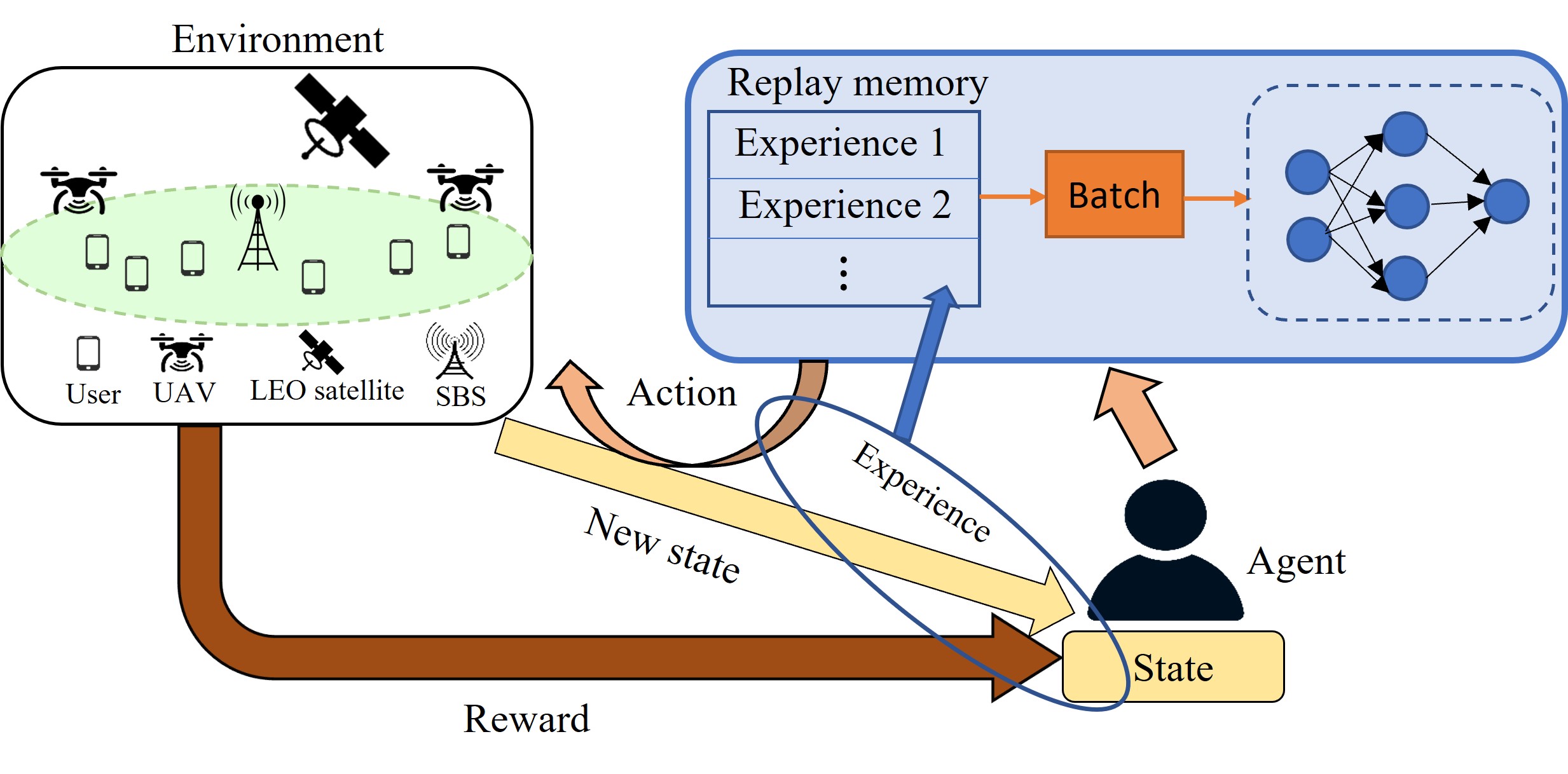}}
\caption{Learning structure based on deep Q-learning algorithm.} \label{DQN_str}
\end{figure}

Among various deep learning algorithms, we review the most commonly used DRL algorithm,  the deep Q-network (DQN) proposed by Mnih et al.  \cite{mnih2015human}. 
{DQN  is a model-free, value-based algorithm and is often considered off-policy due to its use of experience replay.}
DQN uses deep neural networks (DNNs)  to approximate Q-values. It is composed of two neural networks including the evaluation network and the target network \cite{eff_DQN2020}. The current state serves as the input of the evaluation network, and its outputs correspond to the  Q-values evaluated for all actions. The target network provides target values for the evaluation network with the same structure as the evaluation network.  However, there is a delayed update in the weights of the target network to improve 
 stability and reduce the correlation between the Q-values of the evaluation and target networks.  The DNNs are trained by optimizing the loss function

\begin{equation} \label{loss_func}
\begin{aligned}
\begin{split}
l(w) =  E\big[\big(y(t) - Q\big(s(t), a(t); w^{\mathrm{eval}}\big)\big)^2\big],
\end{split}
\end{aligned}
\end{equation}
where $Q(s, a; w^{\mathrm{eval}})$ is the evaluated Q-value from the evaluation Q network with the weights of $w^{\mathrm{eval}}$. Here, $y(t)$ is the target $Q$ value from the target Q network which can be obtained by

\begin{equation} \label{loss_func}
\begin{aligned}
\begin{split}
y(t) = r(t) + \gamma \max_{a^\prime} Q\big(s(t+1), a^\prime; w^{\mathrm{target}}\big),
\end{split}
\end{aligned}
\end{equation}
where $w^{\mathrm{target}}$ is the parameter of the DNN, i.e., weights and
biases. Parameter $\gamma$ denotes a discount factor. 
To train the network, a replay memory with capacity $M_c$ is employed, in which each agent stores its experience $(s_t, a_t, r_t, s_{t+1})$ in the reply memory \cite{Ekram_surv_2021}.
It allows the agent to learn from its earlier memories which contain its current state, action, reward, and next state. To select an action, an $\epsilon$-greedy strategy can be used. Thus, a random action is chosen with a probability $\epsilon$, and the optimal estimate action is selected with probability $1-\epsilon$, as follows:

\begin{equation} \label{loss_func}
\begin{aligned}
\begin{split}
a(t) = 
\begin{cases}
\text{random action}, & \text{with probability } \epsilon \\
\arg\max_a Q(s(t), a; w^{\mathrm{eval}}), & \text{with probability } 1 - \epsilon.
\end{cases}
\end{split}
\end{aligned}
\end{equation}
Algorithm \ref{alg_DQN} presents the DQN approach within a multi-agent system, with the subscript $b$ designating agent $b$.
$N_{ep}$ denotes the total number of episodes or iterations that the DQN algorithm will run. 
This algorithm can be deployed in a distributed manner, in which each agent can have its own DQN, and update it based on the collected data. 
Parameter $N_c$ denotes the delay in updating the weights of the target network. 
Fig. \ref{DQN_str} illustrates the structure of the DQN algorithm. 
%
For the time complexity of the DQN algorithm described in Algorithm  \ref{alg_DQN}, if we consider that step 1 takes $t_0$, steps 6-13 take $t_1$, then the total time taken can be expressed as $t_0 + t_1 \cdot N \cdot |\mathcal{B}| \cdot N_{ep}$. The main term affecting the execution time in this expression is $N \cdot N_{ep}$. Therefore, the time complexity can be derived as $O(N \cdot N_{ep})$.

\begin{algorithm}[t]
\caption{: DQN Algorithm}
\label{alg_DQN}
\begin{algorithmic}[1]
\STATE \textbf{Inputs}: $\mathcal B$,  $M_c$, $N_{ep}, N, \epsilon, \gamma, N_c$
\STATE \textbf{Outputs}: $Q(s_b(t),a_b; w^{\mathrm{eval}}), \forall b \in\mathcal B$ 
\STATE \textbf{Initialization}:  replay memory $M_b$ to capacity $M_c$, $w^{\mathrm{target}}_b$, $w^{\mathrm{eval}}_b$,  let $w^{\mathrm{target}}_b = w^{\mathrm{eval}}_b$, $t=0$, $\forall b\in\mathcal B$
\FOR{ episode = 1 : $N_{ep}$}
\STATE Initialize  $s_t$ for all agents
\WHILE{$t<N$}
\FOR{$\forall b\in\mathcal B$}
\IF{$rand(.) < \epsilon$}
\STATE Select action $a_b(t)$ randomly
\ELSE
\STATE Select action 
{$a_b(t)\!\!=\!\!\argmax_{a_b}\!\! Q(\!s_b(\!t)\!,\!a_b;\!w^{\mathrm{eval}}_b\!)$}
\ENDIF
\STATE Observe reward $r_b(t)$ and transit to state $s_b(t+1)$
\STATE Store the experience $(s_b(t), a_b(t), r_b(t), s_b(t+1))$ in $M_b$
\STATE Sample random minibatch of transitions  $(s_b(j), a_b(j), r_b(j), s_b(j+1))$ 
\STATE Set \\
$y_b(j)\!=\!r_b(j)\!+\!\gamma \max_{a_b^\prime} Q\big(s_b({j+1}), a_b^\prime; w_b^{\mathrm{target}}\big)$
\STATE Perform a gradient descent step on $l_b(w_b^{\mathrm{eval}}) =  E\big[\big(y_b(j) - Q(s_b(j), a_b(j); w_b^{\mathrm{eval}})\big)^2\big]$ with respect to the DQN parameter $w_b^{\mathrm{eval}}$
\STATE Every $N_c$ iterations set $w^{\mathrm{target}}_b = w^{\mathrm{eval}}_b$
\ENDFOR
\ENDWHILE
\ENDFOR
\end{algorithmic}
\end{algorithm}

DQN has been widely adopted in aerial networks to solve different problems such as trajectory design and resource management. In \cite{Aidin19}, the authors optimize the trajectory of a single UAV and scheduling of status update packets. They develop a DQN to minimize the weighted sum of age-of-information. 
To train the UAV,  one fully connected layer with no convolutional neural networks is used.
In \cite{Tang2022}, a double Q-learning-based traffic offloading for SAGINs is proposed. 
In \cite{Wang2020}, a  UAV-aided emergency communications is assumed to overcome the malfunctioning of a ground BS. To maximize the number of users served by the UAV, a DQN-based algorithm is utilized to optimize the UAV's trajectory. 
In \cite{UAV_Mounted2020}, a UAV is used as a mobile edge server to serve users. To optimize the trajectory of the UAV, a QoS-based approach is developed to maximize a reward function that captures the amount of offloaded tasks from users. To solve the problem, a DQN algorithm is developed. To dynamically allocate resources in heterogeneous networks (HetNets) and UAV networks, a DQN-based mechanism is developed in \cite{Kato2020}. The system is composed of a  macro BS and SBSs and a single UAV with considering a high mobility scenario for users. To model the DQN, $4$ hidden layers with different neural units are used. 
In \cite{LuongTWC21}, a joint problem to find the UAVs' locations and manage the resources in a cooperative UAV network is investigated. To solve the problem,  
a DRL-based mechanism combined with a difference of convex algorithm is adopted.
For UAV-enabled wireless powered
communication networks, a joint UAV trajectory planning and resource allocation mechanism is proposed in \cite{WPCN2020}. Accordingly,   a DQL-based strategy is developed to maximize the minimum throughput.

\section{Satisfaction  Based Learning for SAGINs} \label{sec_satis}

The satisfaction concept is proposed in \cite{ross2006satisfaction}, and its applications in the field of wireless communication are introduced in \cite{Tembine10, Perlaza2012_jsel}.
The main difference between RL algorithms and satisfaction-based learning is that the RL algorithms aim at maximizing a reward function, while satisfaction-based learning schemes aim at satisfying the system \cite{meryem_wcnc}. 
Therefore, satisfaction algorithms guarantee that each agent obtains a minimum reward value while improving the total reward of the system.  
{Satisfaction-based learning algorithms can be viewed as a form of policy-based RL, focusing on directly optimizing the agent's policy. Depending on their implementation, they can be either on-policy or off-policy, and they prioritize agent satisfaction as the main optimization objective.} 
Similar to RL algorithms, satisfaction-based learning also requires considering a set of agents and a reward function. Each agent $b$ selects its action according to a  probability distribution $\boldsymbol \pi_b(t) = (\pi_{b,1}(t), \dots, \pi_{b,|\mathcal A_b|}(t))$, where $\mathcal A_b$ is the set of actions for agent $b$. 
Here, $\pi_{b,i}(t)$ denotes the  probability that agent $b$  selects action  $a_{b,i}\in\mathcal A_b$ at time $t$.
In the satisfaction-based approach, the satisfaction means that the observed reward for an agent {is} no less than a certain threshold.
Let $\kappa_b(t)$ denote a satisfaction threshold for agent $b$ at time $t$. In this regard, each learning iteration of the satisfaction approach contains the following process:

\begin{itemize}
\item In the first iteration, each agent $b$ selects its action randomly according to a uniform distribution, i.e. $\pi_{b,i}(t)=\frac{1}{|\mathcal A_b|}$, $\forall b \in \mathcal B$, $\forall a_{b,i} \in \mathcal A_b$ and $i\in\{1, \dots, |\mathcal A_b|\}$.

\item For $t>1$, if agent $b$ is not satisfied with its received reward value, it may change its action based on the probability distribution $\boldsymbol \pi_b(t)$; otherwise, it will keep its current action. Therefore, we define a satisfaction indicator $\varphi_b(t)$ for agent $b$ at time $t$. Let $\Gamma_b(t)$ be the observed reward for agent $b$ at time $t$, each player $b$ computes $\varphi_b(t)$ as follows:

\begin{equation}\label{satis-ind}
\varphi_b(t) =
  \begin{cases}
   1, & \mbox{if } \Gamma_b(t)\geq \kappa_b(t)  \\
   0 , & \mbox{otherwise}.
  \end{cases}
\end{equation}

\item The agent obtains a reward according to its selected action.

\item The probability $\pi_{b,i}(t)$ assigned to action $ a_{b,i} $ is updated as follows:

 \begin{equation} \label{up_pro}
  \pi_{b,i}(t+1)= \begin{cases}
 \pi_{b,i}(t), & \mbox{if } \varphi_b(t) = 1 \\
L_b(\pi_{b,i}(t)), & \mbox{otherwise},
  \end{cases}
\end{equation}
where $L_b(\pi_{b,i}(t))$ is given by 
\begin{equation} \label{satis_eq_up}
\begin{aligned}
\begin{split}
L_b\!\Big(\!\pi_{b,i}(t)\!\Big)\!=
\!\pi_{b,i}(t)\!+\!\mu_b(t)\lambda_b(t)\!\Big(\!\mathbbm{1}_{\{\!a_b(t)=a_{b,i}\!\}}\!-\!\pi_{b,i}(t)\!\Big),
    \end{split}
\end{aligned}
\end{equation}
where $\mu_b(t) =  \frac{1}{1000     t+1}$ and $a_b(t)$ denote the learning rate and the action played by agent $b$ at time $t$, respectively. The parameter $ \lambda_b(t)$ is computed as  $\lambda_b(t) = \frac{\Gamma_{\mathrm{max}} +  \Gamma_b(t) - \kappa_b(t)}{2  \Gamma_{\mathrm{max}}}$, where $\Gamma_{\mathrm{max}}$ is the maximum reward that agent $b$ can achieve.  {However, in the realm of wireless communications,  agent $b$ might not be satisfied at its satisfaction threshold. In this regard, the agent can update the threshold as {$\kappa_{b}(t) \leftarrow \kappa_{b}(t)\cdot(1-\Delta_{\kappa_{b}})$} after each time instant interval {$\tau_{b}$}, where  {$0 < \Delta_{\kappa_{b}} < 1$} is a constant coefficient to decrease the level of the satisfaction threshold.}  
\end{itemize}

{Algorithm {\ref{alg_satis}} presents the pseudocode for the satisfaction mechanism. 
For the time complexity, if we consider step 1 takes $t_0$ and steps 5-13 take $t_1$, the total time taken is $t_0 + N \cdot |\mathcal{B}| \cdot t_1$. The satisfaction-based approach Algorithm \ref{alg_satis} can be derived with the time complexity of $O(N \cdot |\mathcal B|)$.

\begin{algorithm}[tb!]
\caption{: Satisfaction Based Approach}
\label{alg_satis}
\begin{algorithmic}[1]
\STATE \textbf{Inputs}:  $\mathcal B, \mathcal A_b, N, \tau_{b}, \Delta_{\kappa_{b}}$,  $\kappa_b(t)$ for $t=0$,  $\forall b\in\mathcal B$
\STATE \textbf{Outputs}:  $\pi_{b}(t), \varphi_b(t), \forall b \in \mathcal B$
\STATE \textbf{Initialization}:   $t=0$, $\pi_{b,i}(t)=\frac{1}{|\mathcal A_b|}$, $\varphi_b(0)=0$,  $\forall b \in \mathcal B$, $\forall a_{b,i} \in \mathcal A_b$ and $i\in\{1, \dots, |\mathcal A_b|\}$
\WHILE{$t<N$}
\STATE $t \leftarrow t+1$
\FOR {$\forall b \in \mathcal B$}
\IF{$\varphi_b(t-1) = 1$}
\STATE $a_b(t) = a_b(t-1)$
\ELSE
\STATE Select the action 
$a_b(t)$  according to $\boldsymbol\pi_b(t)$
\IF{$\mathrm{mod}(t,\tau_{b}) = 0$}
\STATE $\kappa_b(t)  \leftarrow \kappa_{b}(t)\cdot (1-\Delta_{\kappa_{b}})$
\ENDIF
\ENDIF
\STATE Calculate reward $\Gamma_b(t)$, $\varphi_b(t)$, and $\boldsymbol \pi_b(t)$ 
\ENDFOR
\ENDWHILE
\end{algorithmic}
\end{algorithm}%

 In \cite{Perlaza2012_jsel}, a fundamental concept for satisfaction problems, known as satisfaction equilibrium, is presented, where all agents are satisfied. 
Satisfaction-based learning approaches are utilized in various research areas for resource management in wireless networks. In \cite{Ell_satis2014frequency} and \cite{globe_satis_18}, satisfaction-based approaches are employed for frequency allocation with QoS constraints.
In \cite{perlaza_globecom2010}, a satisfaction-based algorithm is developed to enable a set of transmitters to choose their transmit power levels to ensure a minimum transmission rate.
For sleep mode switching in HetNets, distributed satisfactory schemes are developed in \cite{access_satis_17} and \cite{icee_satis_18}.
Beyond optimizing ground networks through developing satisfactory solutions, the work in \cite{atefeh_PIMRC2020} utilizes a satisfaction mechanism to address the problem of  the $3$D placement of UAVs  integrated into  
TNs to alleviate network overload conditions. 
In addition to UAV placement for ground-integrated networks, satisfaction schemes can be used in SAGINs. In this {framework}, a satisfaction learning-based scheme is investigated in \cite{atefeh_icc2021} for the joint resource management and $3$D UAV trajectory design problem.

\section{PSO for SAGINs} \label{sec_pso}

\begin{figure}[b!]
\centering
{\includegraphics[width=0.8\linewidth]
{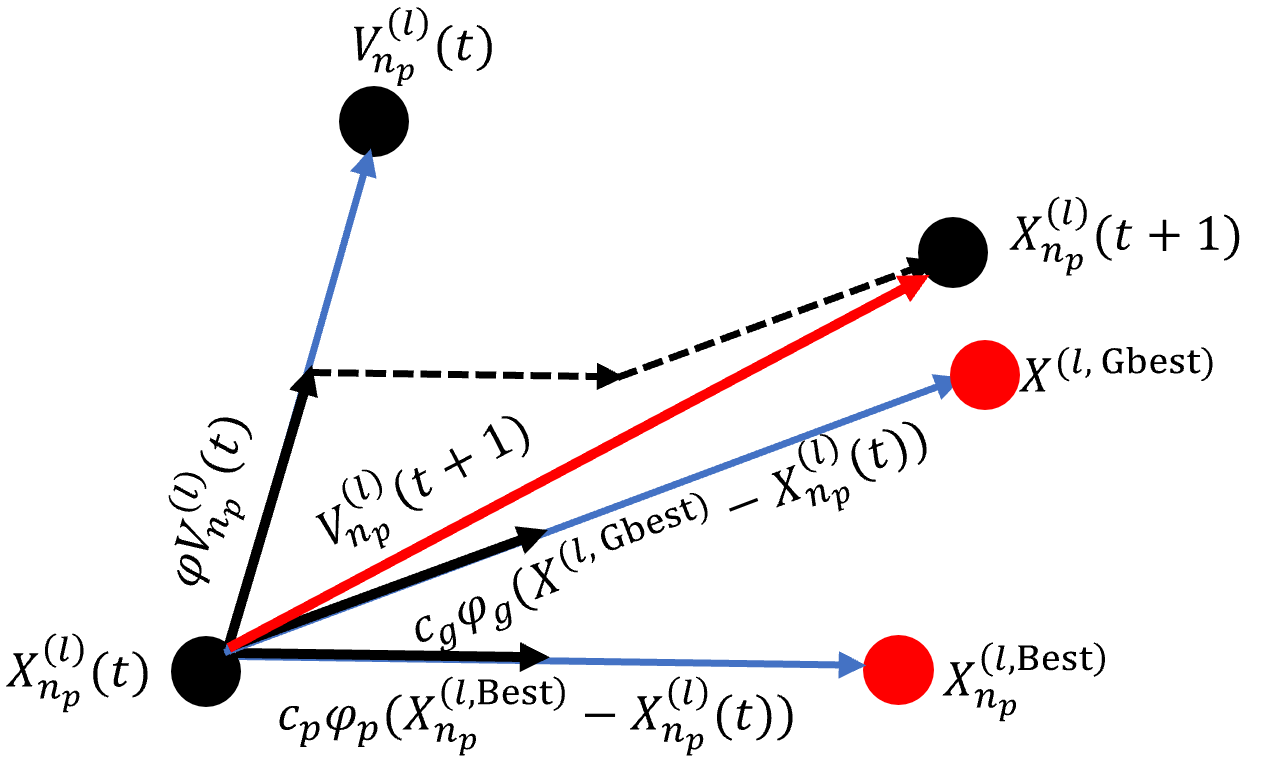}}
\caption{{Illustration for updating the position of an element in PSO algorithm.}} \label{PSO_str}
\end{figure}

\begin{figure}[b!]
\centering{\includegraphics[width=0.95\linewidth]
{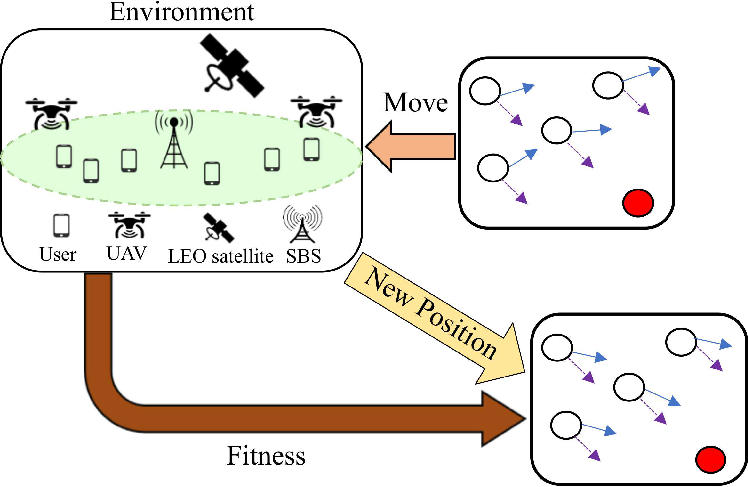}}
\caption{Learning structure based on PSO algorithm.} \label{PSO_str2}
\end{figure}

PSO is a heuristic algorithm based on the concept of population and evolution, inspired by social behavior and movement \cite{kennedy1995particle}. The mechanism of the PSO algorithm is to find the best solution in complicated systems through cooperation and competition between particles in a swarm. The PSO starts with a random set of solutions composed of  $N_p$ particles. 
Each particle $n_p \in N_p$ contains a solution of the problem  {which includes $L$ elements}. The position of each particle is updated according to its velocity. Let $V_{n_p}(t)$ and $X_{n_p(t)}$ denote the velocity and position of particle $n_p$, respectively.
The velocity of an element $l\in L$ in particle  $n_p$ can be updated as follows \cite{atefeh_pso_icee}:

\begin{align} \label{velocity_particle}
 V_{n_p}^{(l)}(t+1)= &\phi V_{n_p}^{(l)}(t)+ c_p \phi_p \big(X_{n_p}^{(l,\mathrm{Best})}-X_{n_p}^{(l)}(t)\big) + \\
 &c_g \phi_g \big(X^{(l,\mathrm{Gbest})}-X_{n_p}^{(l)}(t)\big),   
\end{align}
where $\phi$ denotes  the inertia weight, which is employed to control the impact of the previous history of
velocities on the current particle’s movement.
Parameters $c_p $ and $c_g$ are personal and global learning coefficients, respectively. Here, $\phi_p$ and $\phi_g$ denote two random positive numbers that are uniformly distributed in the interval $[0, 1]$. Therefore, the position of  element $l$ in  particle $n_p$  is updated as follows:

\begin{equation} \label{position_particle}
    X_{n_p}^{(l)}(t+1) = X_{n_p}^{(l)}(t)+ V_{n_p}^{(l)}(t+1). 
\end{equation}
$X_{n_p}^{(l,\mathrm{Best})}$ and {$X^{(l,\mathrm{Gbest})}$ denote the best position of element $l$ in particle $n_p$ and among all particles, respectively. 
Furthermore, we define  the vector $\boldsymbol X_{n_p}(t) = \big(X_{n_p}^1(t), \dots, X_{n_p}^L(t)\big)$,  $\boldsymbol X_{n_p}^{\mathrm{Best}} = \big(X_{n_p}^{(1, \mathrm{Best})}, \dots, X_{n_p}^{(L,  \mathrm{Best})}\big)$, and $\boldsymbol X^{\mathrm{Gbest}} = \big(X^{(1, \mathrm{Gbest})}, \dots, X^{(L, \mathrm{Gbest})}\big)$. 
Let $\Gamma_{n_p}(t)$, $\Gamma_{n_p}^{\mathrm{Best}}$,  $\Gamma^{\mathrm{Gbest}}$ denote the utility function for  particle $n_p$ at time $t$, the best utility of particle $n_p$, and the global best utility, respectively.
}
Fig. \ref{PSO_str} shows an illustration for updating the position of element $l$ in particle $n_p$.
For the time complexity of the PSO algorithm in Algorithm \ref{alg_pso}, if we consider that step 1 takes $t_0$, steps 5-13 take $t_1$, and the particle position update in steps 17-18 takes $t_2$, the total execution time can then be expressed as $t_0 + N \cdot (N_p \cdot t_1  + N_p \cdot |L| \cdot t_2)$. As $t_0$ and $|L|$ are constants, the time complexity of Algorithm  \ref{alg_pso} can be derived as $O(N \cdot N_p )$.

\begin{algorithm}[t]
\caption{: PSO Algorithm}
\label{alg_pso}
\begin{algorithmic}[1]
\STATE \textbf{Inputs}: $N_p, N, L$
\STATE \textbf{Outputs}: $V_{n_p}^{(l)}(t), X_{n_p}^{(l)}(t), \boldsymbol X_{n_p}^{(l,\mathrm{Best})}, \Gamma^{\mathrm{Gbest}}, \Gamma_{n_p}(t), \forall n_p \in N_p$
\STATE \textbf{Initialization}:  
$t = 0$, initialize the position and velocity of all particles, { $\boldsymbol X_{n_p}^{(l,\mathrm{Best})} = X_{n_p}^{(l)}(0)$, $\Gamma^{\mathrm{Gbest}} \leftarrow - \infty$, $\forall n_p \in N_p$ and $l \in L$}

\WHILE{$t<N$}
\STATE $t \leftarrow t+1$
\FOR{$\forall n_p \in N_p$}
\STATE Calculate the reward function 
\IF{$\Gamma_{n_p}(t) > \Gamma_{n_p}^{\mathrm{Best}}$}
\STATE $\boldsymbol X_{n_p}^{\mathrm{Best}} \leftarrow  \boldsymbol X_{n_p}(t)$
\STATE $\Gamma_{n_p}^{\mathrm{Best}}\leftarrow \Gamma_{n_p}(t)$
\ENDIF
\IF{$\Gamma_{n_p}(t) > \Gamma^{\mathrm{Gbest}}$}
\STATE $\boldsymbol X^{\mathrm{Gbest}} \leftarrow \boldsymbol X_{n_p}(t)$
\STATE $\Gamma^{\mathrm{Gbest}}\leftarrow\Gamma_{n_p}(t) $
\ENDIF
\ENDFOR
\FOR{$\forall n_p \in N_p$}
\FOR{$\forall l \in L$}
\STATE Update the velocity $ V_{n_p}^{(l)}(t)$ according to  \eqref{velocity_particle}
\STATE Update the position $X_{n_p}^{(l)}(t)$ according to \eqref{position_particle}
\ENDFOR
\ENDFOR
\ENDWHILE
\end{algorithmic}
\end{algorithm}%

In the context of maximizing throughput and extending coverage for ground users, heuristic algorithms such as PSO can play a key role in solving UAV placement problems \cite{pso_sen19}.   
{The major difference between PSO and RL  algorithms is that PSO benefits from the population-based behavior, and the best solution is shared among the particles in the system,  while the latter aims to maximize the expected cumulative reward for each agent in the population.}
Here, our focus is on research works that utilize PSO methods. 
In \cite{Kalantari_2016}, a PSO-based approach is employed to optimize UAV placement with the objective of minimizing the number of deployed UAVs while ensuring the satisfaction of users' QoS requirements.
Initially, the number of UAVs to serve all users is estimated based on the capacity constraint. 
In the same context,  a PSO-based approach is utilized in \cite{Yuheng_19} to maximize the coverage probability of UAVs by optimizing their $3$D locations. 
In \cite{Zdenek19_access}, the authors address the problem of joint user association and UAV placement using the PSO algorithm to maximize the number of satisfied users in the system. In the first phase,  
 users are associated with the BSs offering the highest signal-to-interference-plus-noise ratio   (SINR). Then, the required bandwidth is allocated to each user to meet their requested data rate until all available bandwidth is assigned. For optimizing the UAV locations, a PSO-based scheme is devised, with the cost function designed to capture user satisfaction based on their minimum required data rates. Based on the new locations of UAVs, the user association and bandwidth allocations are updated.
Another potential application of UAVs is in mobile edge computing networks, where they can serve as flying edge computing servers to offload tasks from users.  In  \cite{cheng_ACM2020}, the authors address the challenge of minimizing energy consumption and delay. They tackle this problem by employing heuristic algorithms such as PSO. 
Table \ref{tab:all_algorithms} summarizes the key characteristics of the described algorithms, including their type, learning approach, action space, and policy method.

\begin{table*}[t!]
  \centering
  \caption{Summary of Learning Algorithms}
  \label{tab:all_algorithms}
\renewcommand{\arraystretch}{1.05}
\begin{tabularx}{\linewidth}{
    >{\hsize=1.83\hsize}X
    >{\hsize=0.83\hsize}X
    >{\hsize=0.83\hsize}X
    >{\hsize=0.83\hsize}X
    >{\hsize=0.83\hsize}X
  }
      \toprule
      Algorithm                      & Type           & Learning Approach  & Action Space      & Policy                    \\ 
      \midrule
      Q-Learning                     & Reinforcement & Model-Free         & Discrete          & Value-Based               \\
      Deep Q-Learning                & Reinforcement & Model-Free         & Discrete/Continuous & Value-Based   \\
      Multi-Armed Bandit (MAB)       & Reinforcement & Model-Free         & Discrete          & Value-Based            \\
      Particle Swarm Optimization    & Optimization  & Heuristic          & Continuous        & N/A                       \\
      Satisfaction-Based Learning    & Satisfaction  &  Model-Free           & Discrete  & Policy-Based    \\         
      \bottomrule
    \end{tabularx}%
\end{table*}

\section{Simulation Results} \label{sec_sim_result}

\begin{table}[t!] 
\vspace{0.2cm}
\begin{center}
\caption{System-Level Simulation Parameters } \label{t_sim_par}
\begin{tabular}{|p{0.8in}|p{1.in}|p{1.in}|} \hline
\multicolumn{3}{|c|}{\textbf{System Parameters}} \\ \hline \hline
\multicolumn{2}{|p{1.8in}|}{\textbf{Parameter}}& \textbf{Value}  \\  \hline
\multicolumn{2}{|p{1.8in}|}{Number of satellites in the orbital plane} & $22$ \\  \hline
\multicolumn{2}{|p{1.8in}|}{Altitude of satellites} & $550$ km\\  \hline
\multicolumn{2}{|p{1.8in}|}{Height of SBSs}  & $15$ m\\  \hline
\multicolumn{2}{|p{1.8in}|}{Height of users}  & $1.5$ m\\  \hline
\multicolumn{2}{|p{1.8in}|}{Maximum altitude of UAVs}  & $121.9$ m\\  \hline
\multicolumn{2}{|p{1.8in}|}{Carrier frequency/channel bandwidth per BS in backhaul links} & $28$ {GHz}, $100$ MHz \\  \hline
\multicolumn{2}{|p{1.8in}|}
{Number of frequency channels}
& $4$ \\  \hline
\multicolumn{2}{|p{1.8in}|}{Channel bandwidth in access link} & $56$ MHz \\  \hline
\multicolumn{2}{|p{1.8in}|}{Noise power spectral density} & $-174$ dBm/Hz \\  \hline
\multicolumn{2}{|p{1.8in}|}{Number of SBSs} & $4$ \\  \hline
\multicolumn{2}{|p{1.8in}|}{Total number of iterations
($N$)} & $5740$\\  \hline
\multicolumn{2}{|p{1.8in}|}{$T_s$} &  $1$ sec\\  \hline
\multicolumn{2}{|p{1.8in}|}{Fixed point iterations} & $500$\\  \hline
\multicolumn{2}{|p{1.8in}|}{$v_{\mathrm{min}}$, $ v_{\mathrm{max}}$}& $0$, $1.3$ m/sec  \\ \hline
  \multicolumn{2}{|p{1.8in}|}{UAV's speed}
& {$10$ m/sec} \\ \hline
\multicolumn{2}{|p{1.8in}|}{$h_{\mathrm{min}}, h_{\mathrm{max}}$} & {$22.5$ m, $121.9$ m} \\ \hline
\multicolumn{2}{|p{1.8in}|}{${\Psi_k} $} & $1.8$ Mbps \\ \hline
\multicolumn{2}{|p{1.8in}|}{$\Phi_{b}, \psi_{b}$} & $0.5, 0.5$ \\ \hline




 \multicolumn{2}{|p{1.8in}|}{Population size for PSO} & $20$ \\ \hline
  \multicolumn{2}{|p{1.8in}|}{Inertia Weight} & $0.9$ \\ \hline
\multicolumn{2}{|p{1.8in}|}{Personal and global learning coefficients} & $0.1, 0.1$ \\ \hline
\multicolumn{2}{|p{1.8in}|}{$\Delta_{\kappa_{b}}, \tau_{b}$} & $0.2, 200$ \\ \hline
\multicolumn{2}{|p{1.8in}|}{Batch size} & $64$ \\ \hline
\multicolumn{2}{|p{1.8in}|}{Replay memory size} & $600$ \\ \hline
\multicolumn{3}{|c|}{\textbf{BS Parameters}} \\ \hline
\textbf{Parameter} & \textbf{Terrestrial BS} & \textbf{UAV} \\ \hline
Transmit power  & $24$ dBm& $24$ dBm  \\ \hline
Reference path loss & LoS: $61.4$ \newline NLoS: $72$ \cite{Rappaport2014}& LoS: $61.4$ \newline NLoS: $61.4$ \cite{hamed_ahmadi2020} \\
\hline
Path loss exponent &  LoS: $2$ \newline NLoS: $2.92$  & LoS: $2$ \newline NLoS: $3$\\ \hline
Shadowing standard deviation  & LoS: $5.8$ \newline NLoS: $8.7$ &  LoS: $5.8$ \newline NLoS: $8.7$ \\
\hline
 \end{tabular}
\end{center}
\vspace*{-0.5cm}
\end{table}

To compare the performance of the PSO-based and learning-based schemes, we consider an area with a size of $500$m $\times$ $500$m. 
A set of users is uniformly distributed within the system and moves based on a random walk mobility model. In this model, users select their speeds from the ranges  $[v_{\mathrm{min}}, v_{\mathrm{max}}]$ and their movement angles from the ranges $[0, 2\pi]$. 
Parameters $v_{\mathrm{min}}$ and $v_{\mathrm{max}}$ indicate the minimum and maximum speed of the users, respectively. Furthermore, $4$ SBSs are uniformly distributed in the system,
 maintaining a minimum distance of $40$ meters from another SBS and $10$ meters from a user. We average our results over $100$ montecarlo simulations. 
To associate the users with the BSs, we consider a policy based on the highest signal strength. The maximum altitude of UAVs is determined based on the Federal Aviation Administration (FAA) Part 107 rules. For backhaul connectivity, we consider a set of LEO satellites that are uniformly distributed in a circular orbit. 
Table \ref{t_sim_par} summarizes the main system parameters used in the simulations. For the performance comparison, we consider the following schemes in our simulations:

\begin{itemize}
    \item \textit{$3$D satisfaction-CA}: Each BS selects its transmission channel and each UAV optimizes its $3$D  trajectory based on the satisfaction approach. 
    \item \textit{$2D$ satisfaction-CA}: The altitude of each UAV is set to the maximum altitude, and the UAVs utilize the satisfaction approach for optimizing their $2$D trajectories.  Furthermore, all the  BSs in the system use the satisfaction approach for channel allocation. 
    \item \textit{$2$D satisfaction}: The altitude of each UAV is set to the maximum altitude, and the $2$D flying directions of the UAVs are optimized based on the satisfaction approach. The channels of all the BSs are selected randomly. 
    \item \textit{$3$D MAB-CA}: The BSs select their transmission channels and the UAVs optimize their $3$D trajectories based on the MAB algorithm.
    \item \textit{$2$D MAB}: The UAVs fly at their maximum altitude and select their $2$D movement direction using the MAB algorithm. Furthermore, all the BSs choose their channels randomly.  
    \item \textit{$3$D PSO}: The PSO algorithm is used to find the $3$D locations of the UAVs. Moreover, the BSs select their channels randomly.

    \item \textit{$2$D PSO}: The altitudes of the UAVs are set to the maximum altitude, and they use the PSO algorithm to find their $2$D positions. Furthermore, all the BSs select their channels randomly.
    
    \item  \textit{Q-learning}: The UAVs fly at the maximum altitude and use the Q-learning algorithm to optimize their $2$D positions.  Moreover, the BSs' channels are randomly selected.  
    
    \item \textit{DQN-CA}: The UAVs select their $3$D trajectories and transmit channels using the DQN algorithm. The DQN uses a fully connected layer with $200$ hidden nodes. 
    

\end{itemize}
 In the following, we provide the performance of the learning and PSO-based algorithms for different number of users and UAVs and address how they affect the BS and user levels performance, i.e., load, fairness, rate, and outage.  Finally, we discuss their performances in the described system.
Here, the solid curves belong to the $3$D trajectories design algorithms, 
and the dashed curves refer to the $2$D trajectories design mechanisms.

\subsection{Performance Evaluation for Various Numbers of Users}

For the first set of results, we consider a system with $4$ SBSs and $4$ UAVs and vary the number of users from $50$ to $300$.

In Fig. \ref{drop-ue}, we compare the outage performance for the PSO and learning-based approaches as a function of the number of users. Outage users are defined as the set of users that receive a data rate less than the requested rate $\Psi_k$.
This performance indicator is affected by the availability of resource at the BSs and the amount of resource to serve the requested rate from the BSs to the users' locations. 
With knowledge of these factors, with increasing the number of users in the system, the remaining resource at the BSs decreases and more users experience the outage conditions. 
It can be observed that the $3$D satisfaction-CA scheme can achieve better outage performance than the other algorithms. Furthermore, for the number of users lower than $250$, the $2$D satisfaction-CA performs better than the $3$D MAB-CA mechanism. 
However, optimizing the altitudes of UAVs is more essential for the high number of users so that  $3$D MAB-CA performs better than the $2$D satisfaction-CA. 
In addition, Fig.  \ref{drop-ue} also demonstrates that the performances of MAB, Q-learning, and PSO-based algorithms for optimizing the $2$D trajectories of UAVs are almost the same while for the high numbers of users, the DQN-CA algorithm yields lower outage users compared to the $2$D based algorithms.

\begin{figure}[tb!]
\hspace*{-0.5cm}
\centering\resizebox{3.7in}{2in}  {\includegraphics
{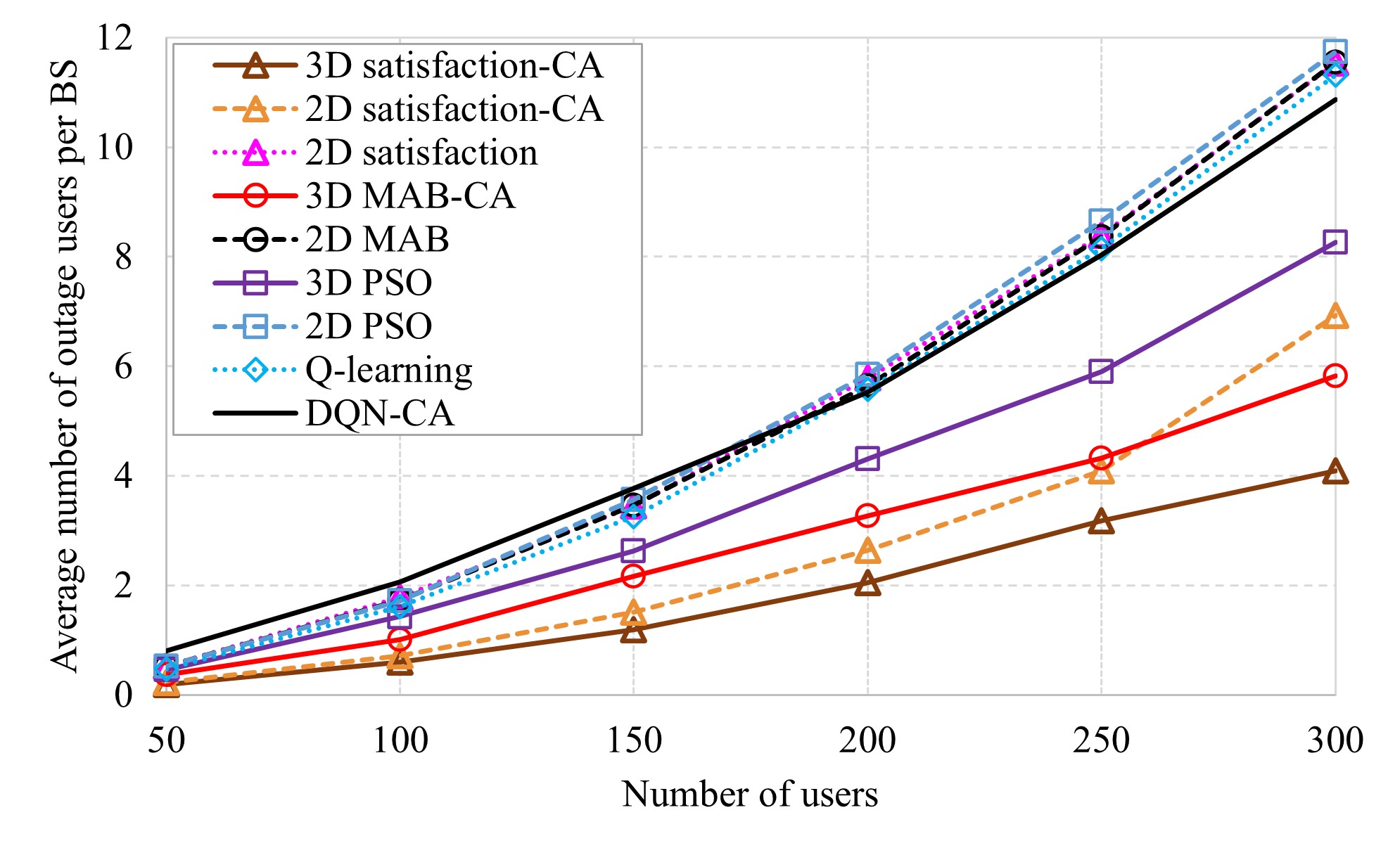}
}
\caption{{Average number of outage users versus the number of users for a system with $4$ SBSs and $4$ UAVs.}} \label{drop-ue}
\end{figure}

In Fig. \ref{load-ue}, we examine the impact of the number of users in the system on the 
average load per BS.  One can observe that the $3$D satisfaction-CA algorithm achieves superior performance in balancing the load among the BSs.
This improvement is primarily due to the enhanced spectral efficiency provided by the $3$D satisfaction-CA. As a result, the average load in the system is reduced, which can be attributed to the inverse relationship between load and rate as described in  \eqref{load_coupled}.
Moreover, as the number of users increases, the average load per BS increases which leads to a rise in the number of outage users. 
However, for a high number of users, the performances of all the sachems are the same and approach to the maximum load, i.e. 1.

\begin{figure}[tb!]
\hspace*{-0.5cm}
\centering\resizebox{3.7in}{2in}  {\includegraphics
{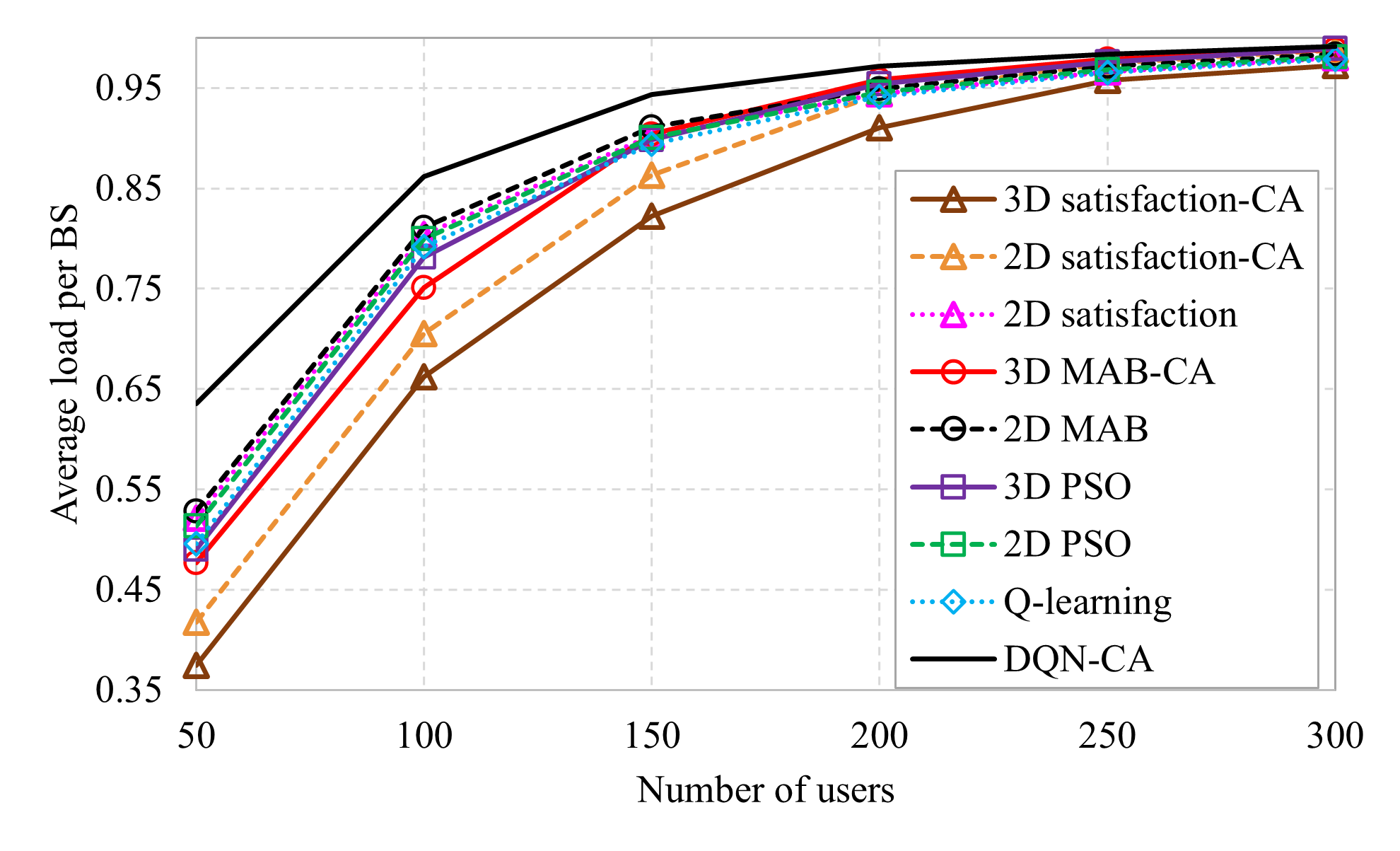}
}
\caption{{Average load per BS versus the number of users for a system with $4$ SBSs and $4$ UAVs.}} \label{load-ue}
\end{figure}

We continue by evaluating the effect of the number of users on the user's rate in Fig. \ref{rate-ue}. It is evident that the data rate is the function of load and the BSs and users' locations. 
An insightful observation from Fig. \ref{rate-ue} is that as the number of users increases, there is a notable decrease in the average rate per user due to the overloading of the BSs (see Fig. \ref{load-ue}). 
The instantaneous rate of a user is directly proportional to the SINR at the user. However, the overall served rate is also related to the fraction of the resources at the BSs. Hence,  highly loaded BSs provide a lower rate over time, such that the long-term service rate experienced by the users is related to the load of the system.
Furthermore, under light load conditions,  users experience better SINR compared to high load conditions due to lower interference. 
On the other hand, the average rate also depends on the resource management scheme at the BSs.
Accordingly, the $3$D satisfaction-CA approach significantly improves the average rate per user compared to the other approaches due to its load-balancing mechanism and efficient resource management.

\begin{figure}[tb!]
\hspace*{-0.5cm}
\centering\resizebox{3.7in}{2.2in}  {\includegraphics
{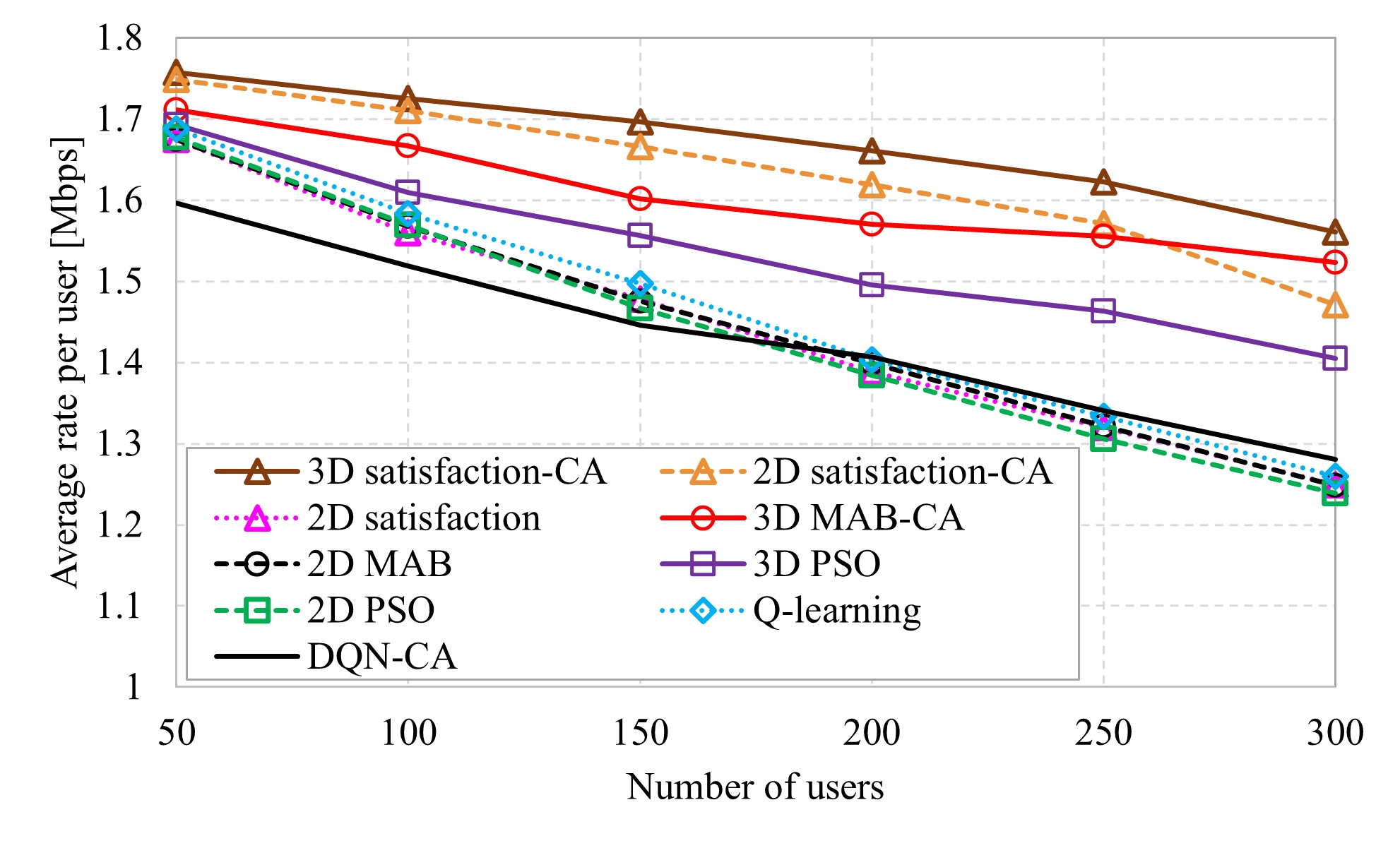}
}
\caption{{Average rate per user versus the number of users for a system with $4$ SBSs and $4$ UAVs.}} \label{rate-ue}
\end{figure}

In Fig. \ref{fair-ue}, we compare the average fairness measure as given by the index that we have introduced in \eqref{fairness_metric} for different numbers of users.  It shows that the average fairness among all the schemes
demonstrates a similar trend,  with a decrease observed as the number of users increases. 
This decline can be attributed to the potential inadequacy of resources available for allocation to users, especially as their numbers grow.
Furthermore, the $3$D satisfaction-CA approach outperforms the other methods in terms of fairness.

\begin{figure}[tb!]
\hspace*{-0.5cm}
\centering\resizebox{3.7in}{2.2in}  {\includegraphics
{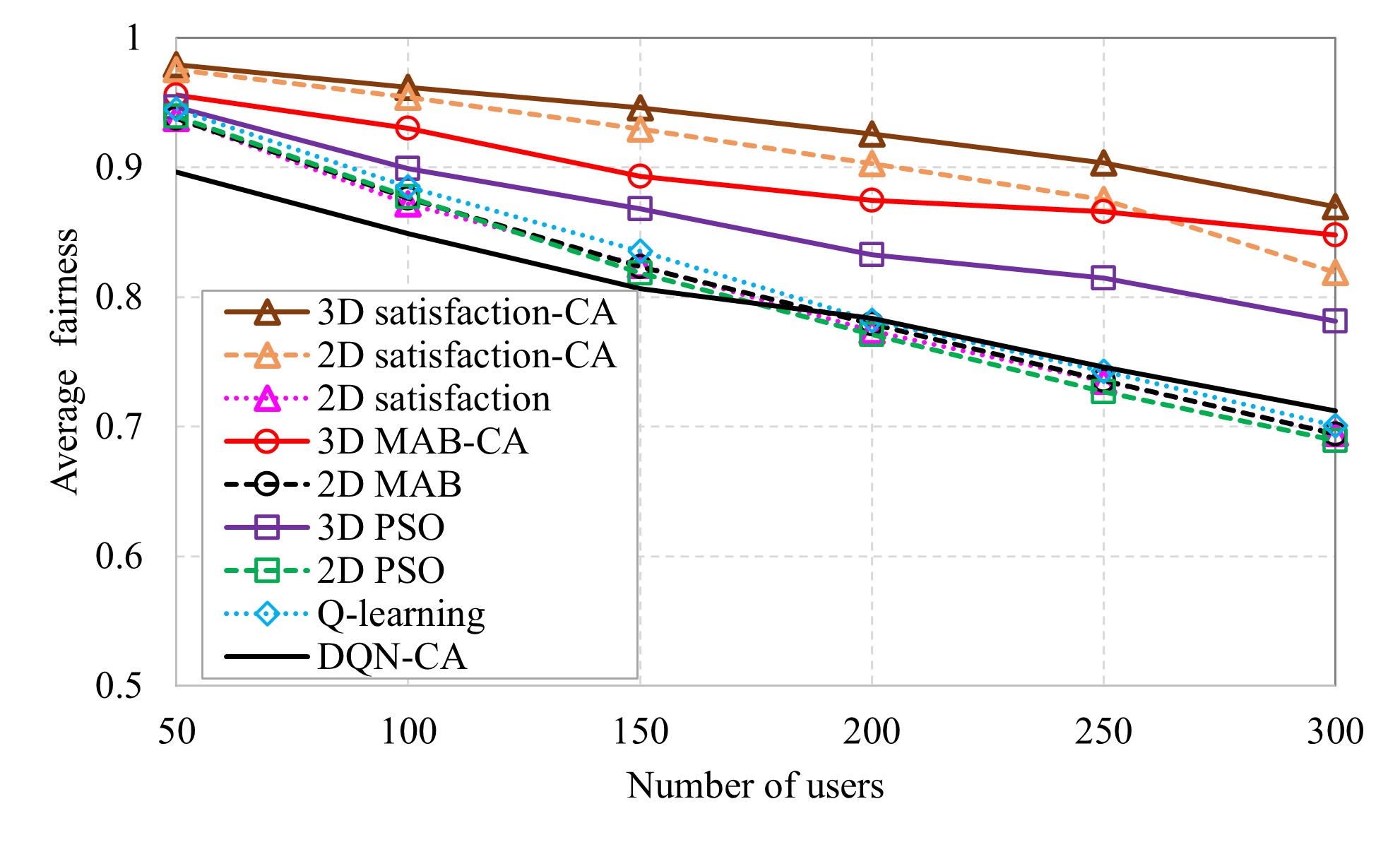}
}
\caption{{Average fairness versus the number of users for a system with $4$ SBSs and $4$ UAVs.}} \label{fair-ue}
\end{figure}

Fig. \ref{reward-ue} shows the behavior of the reward function defined in \eqref{reward_func}. 
  As the number of users increases, the $3$D satisfaction-CA scheme demonstrates superior performance in terms of average fairness and load, resulting in better average rewards compared to other approaches. We can observe that for the high number of users, the performances of the $2$D MAB, PSO, Q-learning, and DQN algorithms are almost the same.

\begin{figure}[tb!]
\hspace*{-0.5cm}
\centering\resizebox{3.7in}{2.2in}  {\includegraphics
{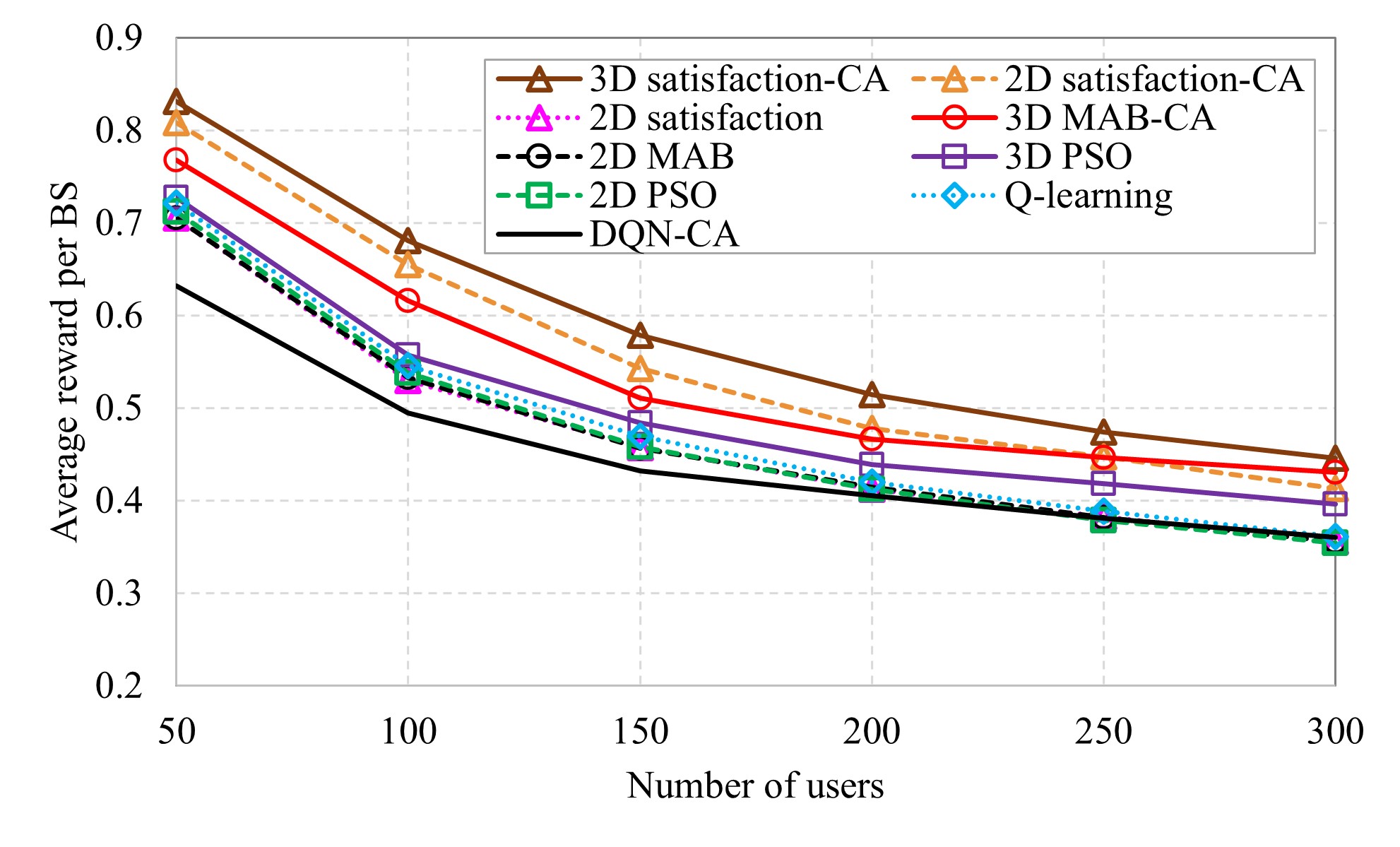}
}
\caption{{Average reward versus the number of users for a system with $4$ SBSs and $4$ UAVs.}} \label{reward-ue}
\end{figure}

\subsection{Performance Evaluation for Various Numbers of UAVs}

Here, we vary the number of UAVs in the system and observe the performance of the learning and PSO mechanisms. Moreover, we set the number of users and the number of SBS to $150$ and $4$, respectively. 

Fig. \ref{drop-uav} illustrates the average number of outage users per BS against varying numbers of UAVs. It shows that offloading the users from the highly loaded BSs to the lightly loaded BSs through increasing the number of UAVs helps to improve the average rate per user and decrease the number of outage users.
 Furthermore, Fig. \ref{drop-uav} reveals that 
dense deployments of UAVs result in comparable performances between the $3$D-based mechanisms and the $2$D satisfaction-CA scheme, with both approaches approaching approximately one outage user per BS.

\begin{figure}[tb!]
\hspace*{-0.5cm}
\centering\resizebox{3.7in}{2.2in}  {\includegraphics
{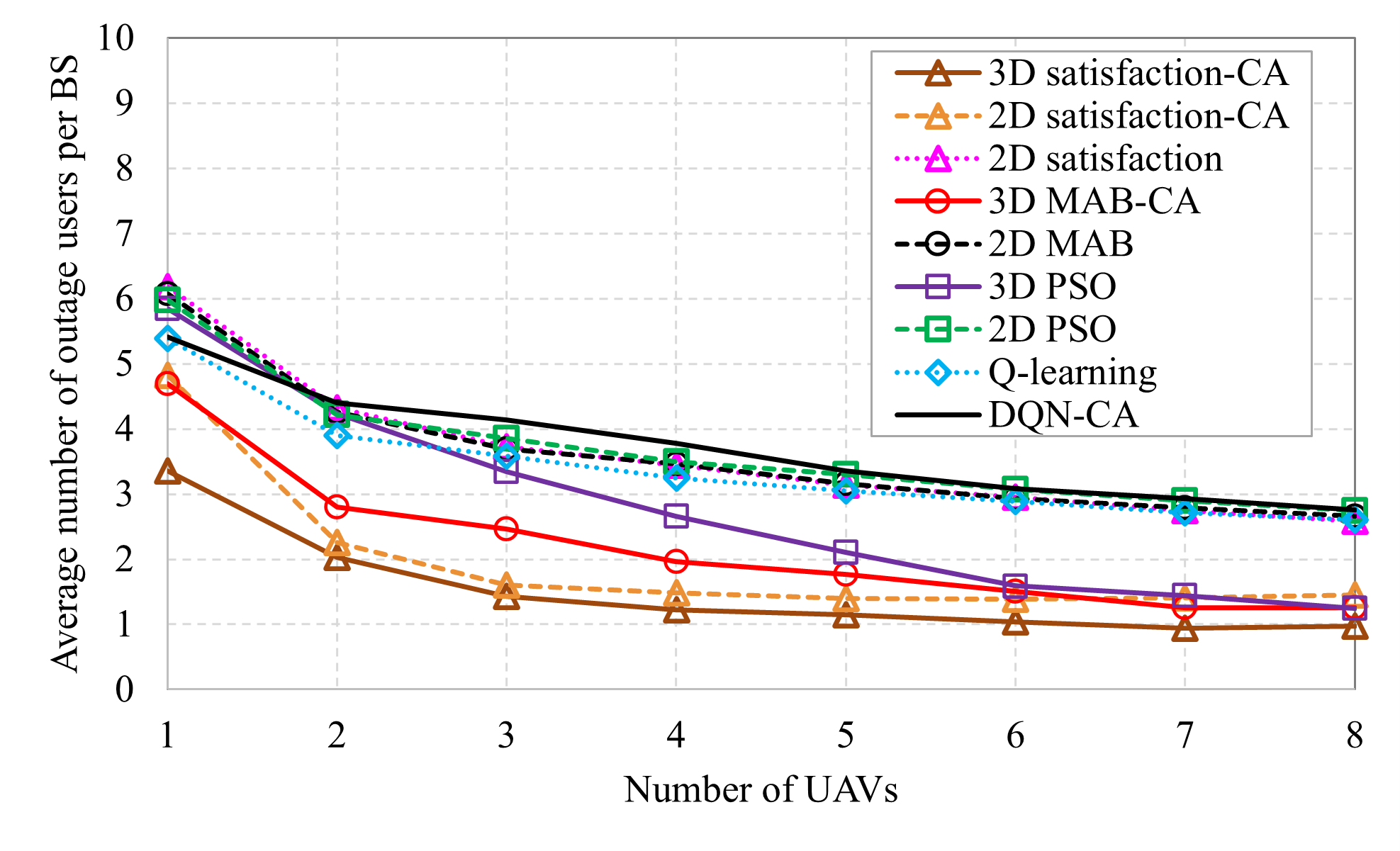}
}
\caption{{Average number of outage users per BS versus the number of UAVs for a system with $4$ SBSs and $150$ users.}} \label{drop-uav}
\end{figure}

Fig. \ref{load-uav} demonstrates how increasing the number of UAVs impacts load balancing within the system. 
For the dense deployment of UAVs,   
there is a decrease in the average load per BS. This reduction in load is due to the offloading of load from highly loaded BSs to lightly loaded ones by employing additional UAVs. Furthermore, we can see that for scenarios with fewer than $7$ UAVs, the $3$D satisfaction-CA approach outperforms the others. However, with a higher number of UAVs, the $3$D PSO scheme demonstrates efficient load balancing. 
{This is due to that PSO is an algorithm that uses a swarm of particles to explore the solution space.  By optimizing the locations of UAVs in the $3$D space, the PSO algorithm 
 effectively minimizes interference and maximizes SINR, even when dealing with a large number of UAVs.
On the other hand, increasing the number of UAVs may increase the interference in the system, consequently leading to increased load. This trend is observable in the $2$D satisfaction-CA approach when the number of UAVs exceeds $6$ which causes a rise in the average load per BS.


\begin{figure}[tb!]
\hspace*{-0.5cm}
\centering\resizebox{3.7in}{2in}  {\includegraphics
{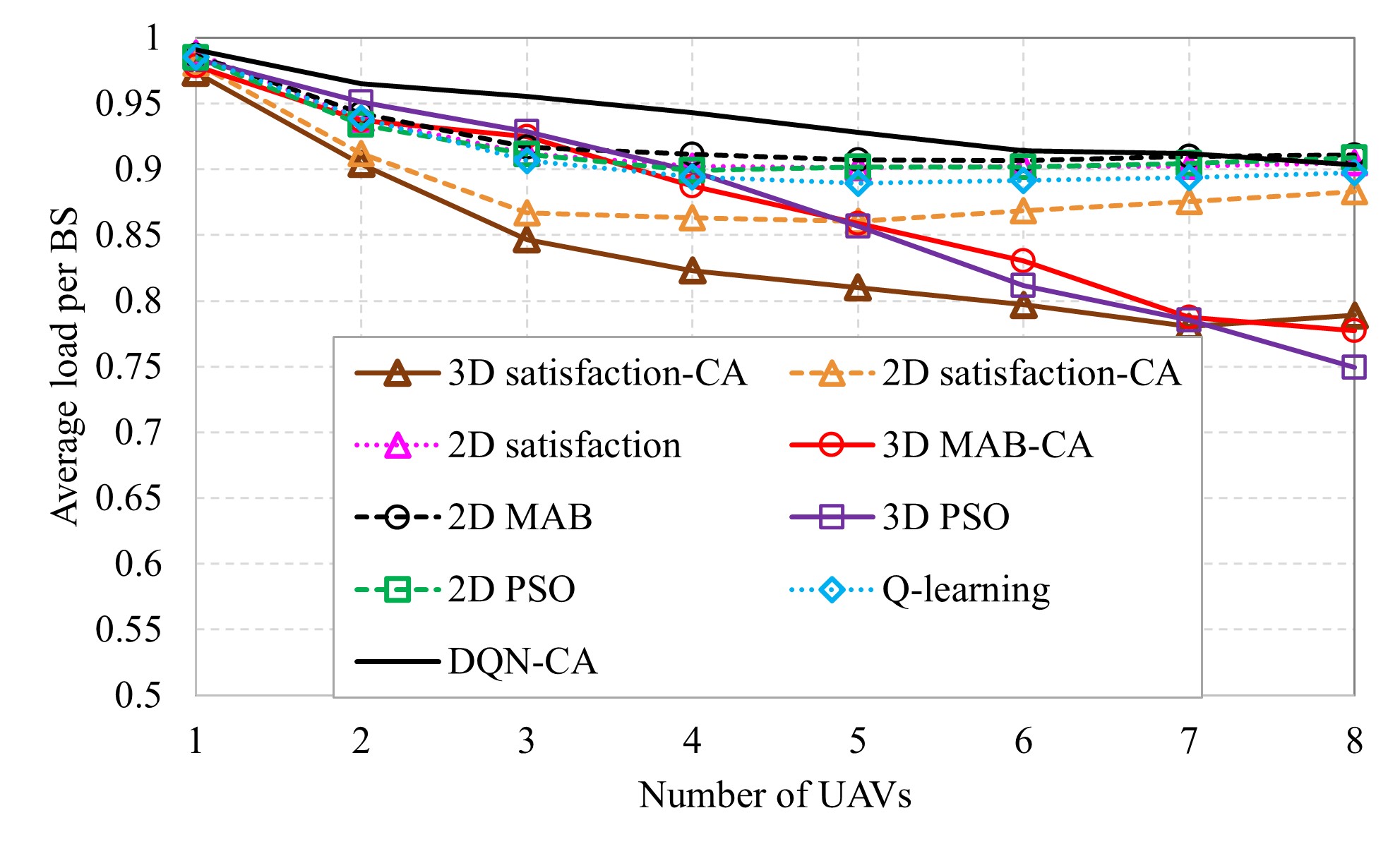}
}
\caption{{Average load per BS versus the number of UAVs for a system with $4$ SBSs and $150$ users.}} \label{load-uav}
\end{figure}

In Fig. \ref{rate-uav}, we compare the performance of the PSO and learning-based schemes in terms of the average rate per user. It can be observed that the $3$D satisfaction-CA approach outperforms the other approaches. Moreover, the $2$D based algorithms, except for $2$D satisfaction-CA and DQN-CA, have almost the same performance. Furthermore, with an increase in the number of UAVs in the system, the users have more opportunities to associate with lightly loaded BSs, thereby improving their rates.  However, as we mentioned earlier, this increase in the number of UAVs may also lead to higher interference in the system.

\begin{figure}[tb!]
\hspace*{-0.5cm}
\centering\resizebox{3.7in}{2in}  {\includegraphics
{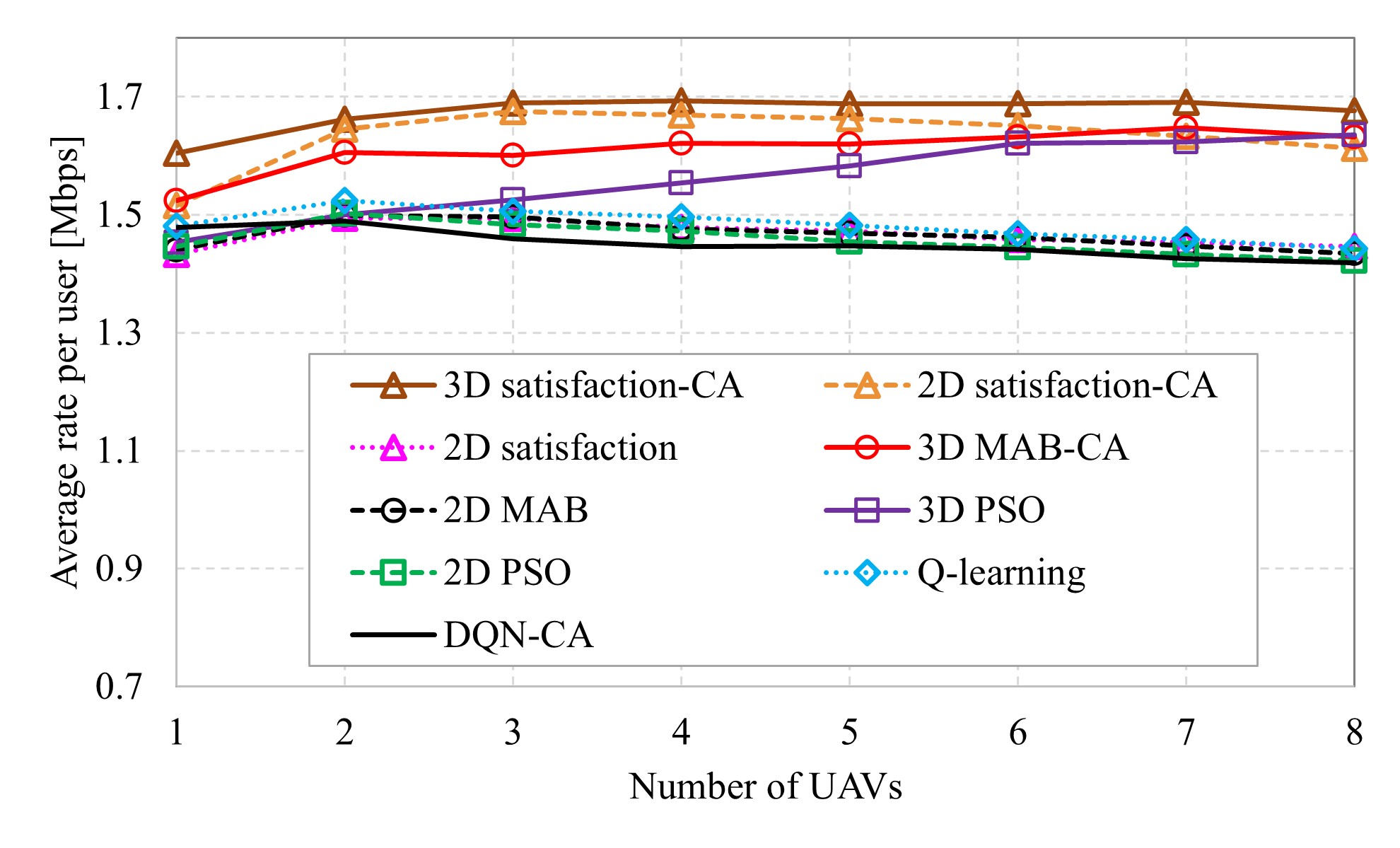}
}
\caption{{Average rate per user versus the number of UAVs for a system with $4$ SBSs and $150$ users.}} \label{rate-uav}
\end{figure}

The improvement in the average fairness depicted in Fig. \ref{fair-uav}
correlates with the findings presented 
in Fig. \ref{rate-uav}. 
As shown in Fig. \ref{fair-uav}, increasing the number of UAVs leads to enhanced average fairness due to the increase of available resources.
Indeed, the increased number of UAVs corresponds to a broader coverage of users. 
However, the performance can be impacted by the increasing co-channel interference in the system.  Furthermore, it can be observed that the $3$D satisfaction-CA algorithm achieves the highest fairness for all the number of UAVs, while the $2$D MAB and $2$D PSO, Q-learning, and DQN-CA have inferior fairness.
This is because the $3$D  satisfaction-CA algorithm can manage increasing interference through effective resource allocation across all the BSs and optimized trajectory designs for the UAVs.  Consequently, the performance gap between the $3$D satisfaction-CA algorithm and the $2$D MAB, PSO, Q-learning,  and DQN-based algorithms increases. Furthermore,  
when the number of UAVs exceeds   $7$, both the $3$D MAB-CA and $3$D PSO algorithms approach the performance level of the  $2$D satisfaction-CA algorithm and show slight improvements in the fairness index.  

\begin{figure}[tb!]
\hspace*{-0.5cm}
\centering\resizebox{3.7in}{2in}  {\includegraphics
{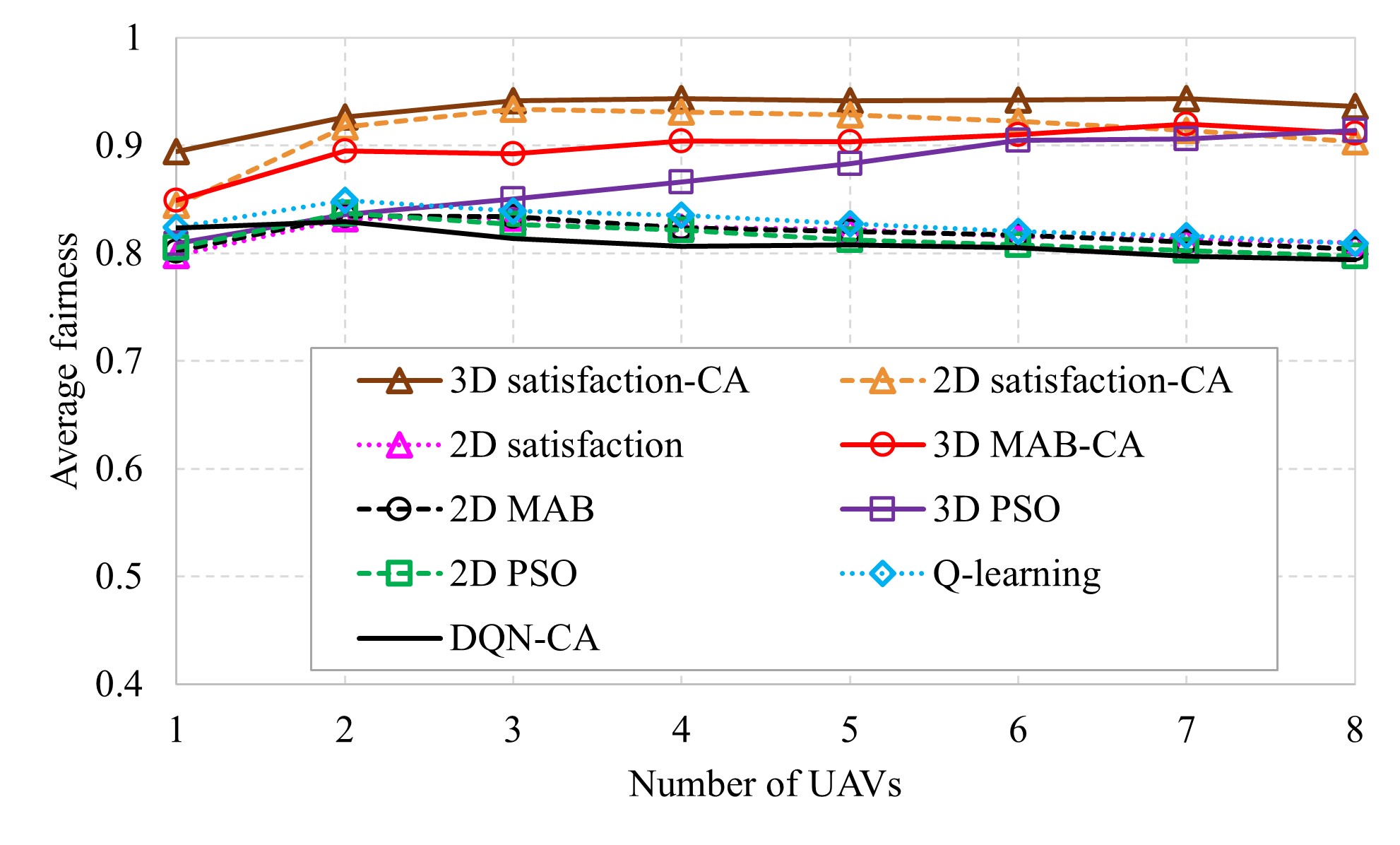}
}
\caption{{Average fairness versus the number of UAVs for a system with $4$ SBSs and $150$ users.}} \label{fair-uav}
\end{figure}

Fig. \ref{reward-uav} shows the enhancement in the average reward per BS 
 as the number of UAVs increases.
It is worth
noting that the performance of the $3$D satisfaction-CA approach outperforms the other methods due to its efficient optimization of channel allocation and $3$D trajectories of the UAVs. The DQN-CA and $2$D based approaches, including $2$D satisfaction, $2$D PSO, $2$D MAB, and Q-learning have the lowest reward values compared to the other methods. 
 Furthermore, 
 for fewer than $6$ UAVs,
 the $2$D satisfaction-CA approach outperforms the others, except for the $3$D satisfaction-CA algorithm. However,   as the number of UAVs increases, the $3$D MAB-CA and $3$D PSO approaches yield higher reward values.   
 This behavior
 results from their performances being
proportional to the load and fairness values, as illustrated in Fig. \ref{load-uav} and Fig. \ref{fair-uav}.

\begin{figure}[tb!]
\hspace*{-0.5cm}
\centering\resizebox{3.7in}{2in}  {\includegraphics
{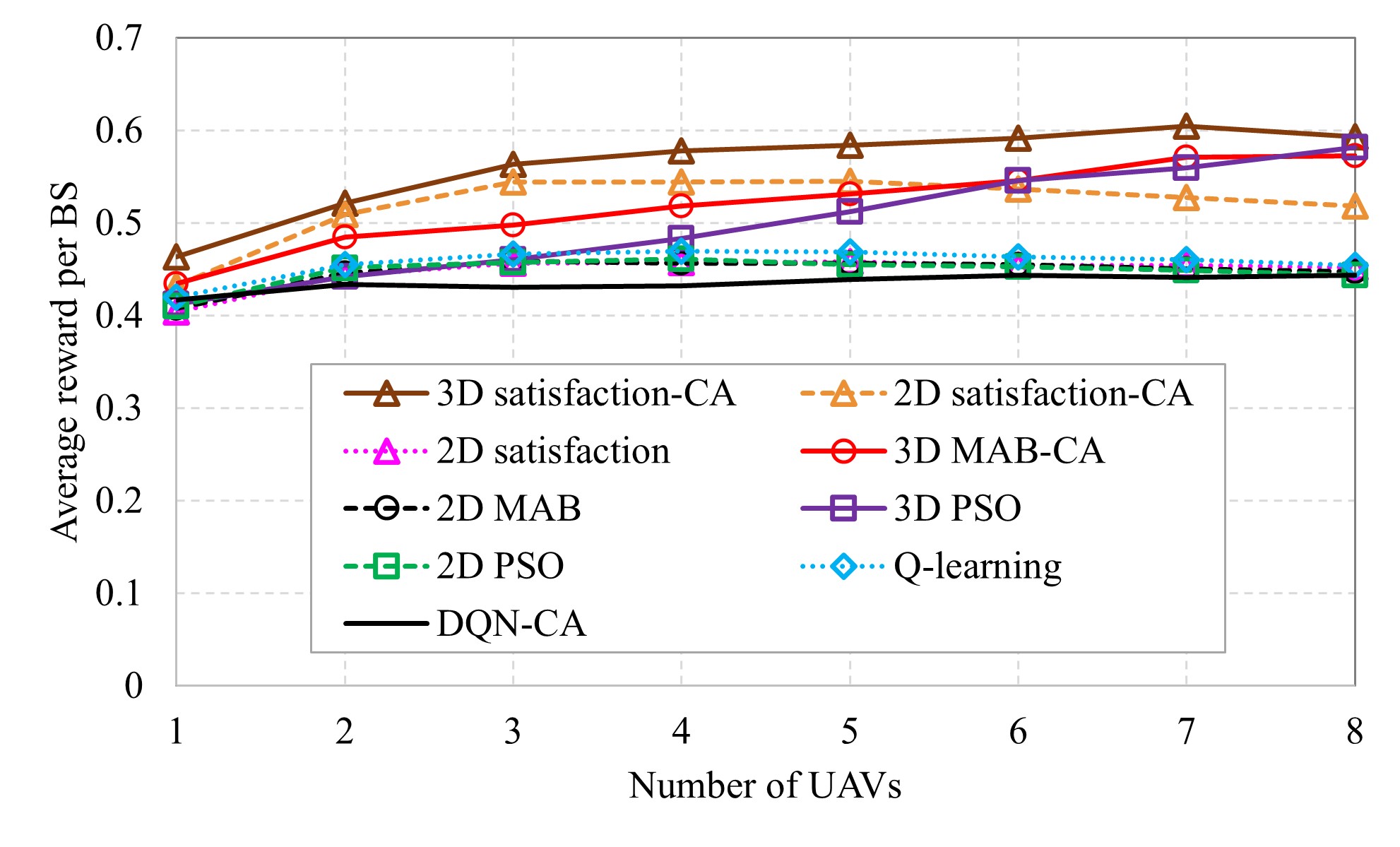}
}
\caption{{Average reward versus the number of UAVs for a system with $4$ SBSs and $150$ users.}} \label{reward-uav}
\end{figure}


\begin{figure}[tb!]
\hspace*{-0.5cm}
\centering\resizebox{3.7in}{2in}  {\includegraphics
{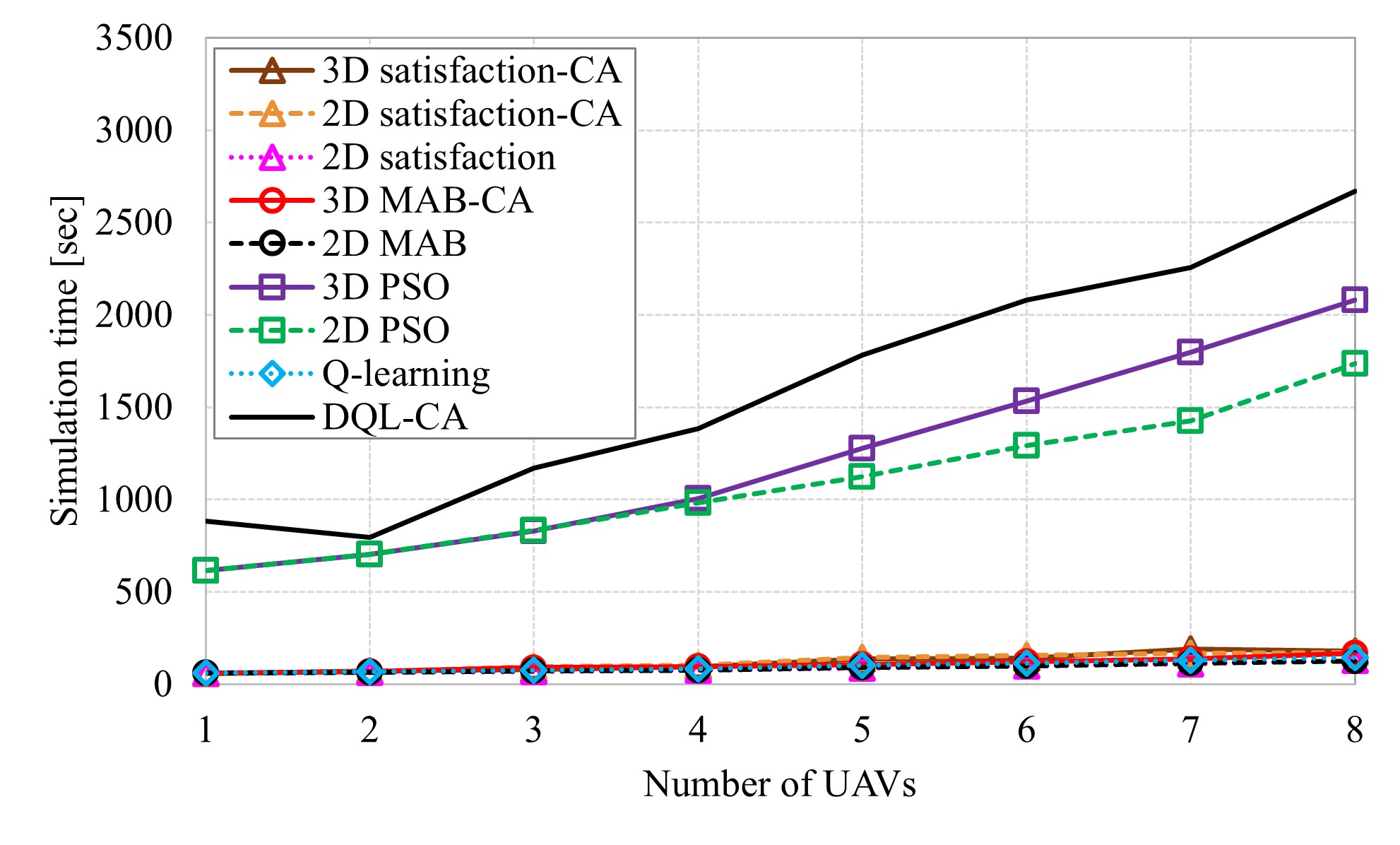}
}
\caption{{Simulation time versus the number of UAVs for a system with $4$ SBSs and $150$ users.}} \label{time_alluav}
\end{figure}

In Fig. \ref{time_alluav}, we evaluate the performance of the learning and PSO-based algorithms in terms of time consumption for the simulations. We can observe that the DQN-CA and PSO-based algorithms consume more time compared to the other algorithms 
due to the involvement of several neural networks and particles.
Furthermore, satisfaction-based, MAB-based, and Q-learning algorithms have the lowest simulation times. The results are aligned with our time complexity analysis.
In addition, the dimension of UAV trajectories (e.g., $2$D or $3$D trajectories) contributes to the total execution time of an algorithm as a constant factor. The complexity upper bounds described in Sections \ref{sec_RL_SAGIN} to \ref{sec_pso} remain the same. This is evidenced by the simulation times logged from Fig. \ref{time_alluav}.

\subsection{Convergence Behaviour}
{To show the convergence of the proposed $3$D trajectory mechanisms outlined in this paper, we conducted simulations to evaluate the convergence behavior and validate their performance. Specifically, we performed simulations for the $3$D MAB-CA, $3$D satisfaction-CA, DQN-CN, and $3$D PSO mechanisms for a system with $4$ UAVs, $4$ SBSs, and $300$ users.
As shown in Fig. {\ref{3d_conv}}}, {the $3$D MAB-CA algorithm demonstrates rapid convergence which reaches a stable solution within approximately $100$ iterations.  This rapid convergence makes the $3$D MAB-CA approach suitable for dynamic environments where quick adaptation is crucial.
{The $3$D satisfaction-CA algorithm converges after about $800$ iterations. This mechanism prioritizes ensuring each agent's satisfaction, which involves meeting specific thresholds before optimizing further. This results in a more gradual convergence compared to the $3$D MAB-CA approach but still achieves a stable solution within a reasonable number of iterations. This makes it a robust choice for scenarios requiring reliable performance guarantees. The DQN approach, as depicted in {Fig. \ref{3d_conv}}}, {converges after approximately 1200 iterations. The DQN algorithm leverages neural networks to approximate the value function, which requires more iterations to train effectively. Finally, 
{the $3$D PSO algorithm also converges after about $1200$ iterations.  While it can handle continuous action spaces, it generally requires more iterations to fine-tune the solution.}
\hl{It is important to highlight that for scenarios with fewer than $300$ users, the $3$D Satisfaction-CA exhibits even better performance, significantly outperforming the $3$D MAB-CA. For higher numbers of users, the performance of the $3$D Satisfaction-CA approaches that of the $3$D MAB-CA, indicating that both algorithms perform comparably well in high-density scenarios. While it is true that the convergence iteration number of the $3$D Satisfaction-CA is significantly larger than that of the $3$D MAB-CA, this additional convergence time is a trade-off for substantial gains in fairness, load balancing, and overall system performance, especially in scenarios with fewer users.}

\begin{figure}[tb!]
\centering\resizebox{3.5in}{3in}  {\includegraphics
{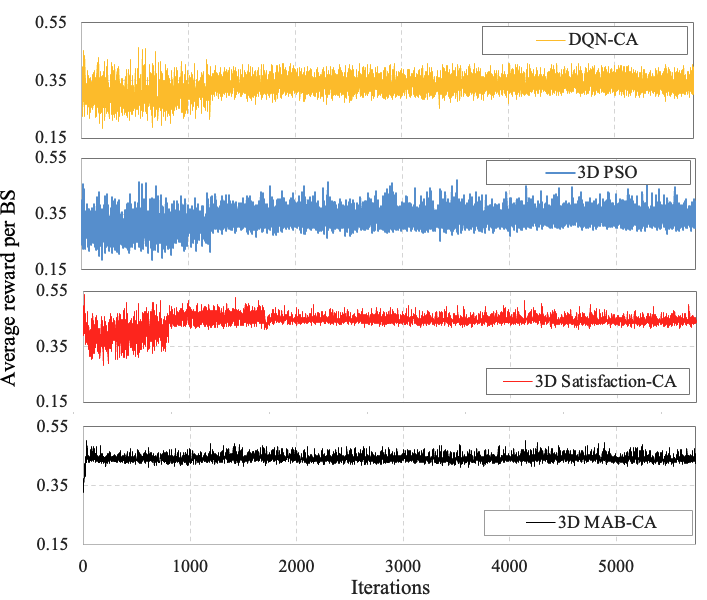}
}
\caption{{The convergence behavior  of the $3$D approaches for a system  with $4$ UAVs, $4$ SBSs and $300$ users.}} \label{3d_conv}
\end{figure}





\subsection{Results Discussion}

In this work, we address the channel allocation problem combined with the trajectory design for the UAVs. Simulation results indicate that the $3$D satisfaction-CA approach outperforms alternative methods. However,  
selecting appropriate parameter values for the satisfaction-based algorithm, such as the satisfaction threshold, is crucial for its effective implementation.
By incorporating the channel allocation mechanism, the $3$D satisfaction-CA approach enhances system performance compared to other $3$D-based approaches.
In addition, the satisfaction-based approaches avoid randomly changing actions when an agent is satisfied, which results in better performance.
Therefore, efficiently allocating resources is crucial. 
To address this, we employed both RL and PSO algorithms for optimizing SAGINs. While both algorithms are powerful, our detailed comparison revealed several crucial differences. RL, inspired by behavioral psychology, adapts strategies based on real-time feedback, making it highly adaptable to dynamic environments. RL agents learn from past experiences, adjusting their behavior over time, a capability lacking in PSO particles. Furthermore, RL excels in high-dimensional spaces, essential for modeling complex SAGIN scenarios. In our SAGIN study, RL's adaptability, learning ability, and efficiency in high-dimensional spaces were paramount. Its dynamic adjustments in both trajectory and channel allocation based on real-time feedback and environmental changes were key to outperforming PSO.}

In addition, while we optimize the $2$D locations of the UAVs, determining optimal altitudes for the UAVs is an important issue. In this regard, we can observe that the $3$D based algorithms outperform the $2$D based for the high numbers of UAVs. However, the performance of the DQN-CA algorithm is lower than the other $ 3$D-based algorithm. The DQN structure used in this paper is similar to \cite{Aidin19} for a system with only one BS which is a UAV. We use this structure for a more complex system which is our initial attempt to develop DQN into SAGINs. For our future
work, we will continue to consider novel and more complex DQN algorithms and improve the computation overhead of the algorithm.
Regarding the simulation time, the DQN-CA and PSO-based mechanisms consume more time compared to the other approaches due to having two neural networks and population-based properties, respectively. Furthermore, with increasing the number of agents in the system, the simulation time for the learning-based and PSO algorithms increases. However, as a representative metaheuristic algorithm, the PSO-based approach can have a marginal performance difference compared to RL algorithmic approaches when the number of UAVs is small. Due to its problem-independence feature, PSO can be easily transferred to other UAV-assisted SAGIN scenarios. The convergence time taken in simulation can be further reduced with some schemes, such as random sampling of control parameters \cite{Sun2019}.

\hl{The performance metrics illustrated in our simulations such as the average number of outage users, average load, average rate,  and average fairness are all directly related to the total system reward function. The reward function we employ incorporates both fairness and load. The \textit{average number of outage users} reflects the system's ability to maintain connectivity and provide consistent service, which indicates effective resource management, enhances fairness, reduces load, and improves overall system performance. The \textit{average load} on each BS, a direct component of our reward function, signifies efficient resource utilization and contributes to higher user satisfaction. The  \textit{average rate} reflects the quality of the connection each user experiences, influenced by load distribution and interference management. Finally, \textit{average fairness} ensures equitable resource distribution, maintaining user satisfaction and preventing service degradation, leading to higher reward values and optimizing the overall reward function. }

\subsection{Open Challenges}

Our work has provided a solid performance benchmark for recent algorithmic approaches for UAV-assisted SAGINs. However, there are still some open challenges for future contributions. The following are examples from the perspectives of system architectures, application scenarios, and algorithmic variations. 

First of all, the detailed modeling of UAV networking, where UAVs can collaborate for message exchanges can be extended from our baseline system model. We realized that the networking functionality between UAVs can be optional and the assumption of system capability on UAV hardware and software is often required. For this reason, multi-hopping is not considered in our formulation. However, for wide area coverage using many UAVs, multi-hopping, and networking may be advantageous for UAV fleet operations, and the networking-focused technical analysis needs additional work. 

Second,  variations in UAV roles within a SAGIN can lead to the extended system model for real-world scenarios. Although we focus on the role of UAVs as aerial BSs, they can also be modeled as relay nodes \cite{Gapeyenko20}. Additionally, the consideration of variable speed and altitude-keeping schemes, as well as complex cooperative schemes for state updates, require further exploration.

Third, setting the objective functions and constraints in the algorithms to support additional application-specific scenarios is another challenge. We have considered as many factors as possible to mimic a real-world scenario. However, some unexpected situations may still exist, such as weather and environmental conditions affecting link conditions, which can be hard to predict and mathematically model. This still requires future work.

Last but not least, the algorithmic variations based on the algorithms we presented in this paper may be employed to further improve the performance metrics. For example, a DRL method \cite{Mnih2015} could be used to efficiently handle dynamics and scalable problems. Deep Q-learning algorithms can address state-space explosions and enhance efficiency for complex optimization problems. Additionally, PSO has recently been adopted in ensemble methods for ML algorithms, such as neural networks, for hyper-parameter tuning. It has also been utilized as a heuristic search scheme for RL algorithms. Another future direction can be the combined use of PSO and RL to manage more dynamic scenarios arising from our formulated problem.

\section{Conclusion} \label{sec_conclusion}
Integrating UAVs into space networks and TNs faces many challenges in terms of complexity, scalability, and flexibility requirements for the next-generation telecommunications networks. Although a UAV-assisted SAGIN is promising and adaptable to various scenarios, deployment considerations are often viewed as barriers to real-life applications. This paper has reviewed the recent algorithmic approaches in RL, satisfaction-based learning, and heuristic methods. We provide key technical comparisons between these approaches in real-world scenarios. Our evaluation results have provided simulation outcomes to compare the performance metrics among the representative algorithms in these approaches. The results reveal that the satisfaction algorithm combined with channel allocation outperforms the other algorithms. With fair, consistent, and technical comparisons, our work can guide the future design of UAV-assisted SAGIN missions and systems, including SAGIN-based Internet of Things applications.

{While our current work does not include simulations of DDPG and PPO, we acknowledge the significance of these advanced RL techniques in the domain of SAGINs. DDPG and PPO offer promising solutions for addressing challenges related to adaptive decision-making, resource optimization, and network management in dynamic and heterogeneous environments. Extending the current setting to continuous action spaces and incorporating a comparison of these approaches will be left for future work.}

\section*{Acknowledgment} 
This work was supported by the High-Throughput and
Secure Networks Challenge program of National Research Council Canada under Grant No. CH-HTSN-418. We also acknowledge the support of the Natural Sciences and Engineering Research Council of Canada (NSERC), [funding reference number RGPIN-2022-03364].

\bibliographystyle{IEEEtran}
\bibliography{IEEEabrv,./main.bib}

\end{document}